\documentclass
[aps,prc,nofootinbib,
preprint
]{revtex4-1}
\usepackage{graphicx,subfigure,latexsym}
\usepackage{amsmath,amssymb,amsfonts}
\usepackage{ulem}
\usepackage{color}
\usepackage{bm}
\usepackage{url,hyperref}

\newcommand{\be}{\begin{equation}}
\newcommand{\ee}{\end{equation}}
\newcommand{\bear}{\begin{eqnarray}}
\newcommand{\eear}{\end{eqnarray}}
\newcommand{\ba}{\begin{array}}
\newcommand{\ea}{\end{array}}

\def \be {\begin{equation}}
\def \ee {\end{equation}}

\def \bes {\begin{subequations}}
\def \ees {\end{subequations}}

\def \pd {\partial}

\def \eq {Eq.~}

\def \<{\langle}
\def \>{\rangle}
\def \+{\dagger}
\def \({\left(}
\def \){\right)}
\def \[{\left[}
\def \]{\right]}

\def \vB {\bm{B}}

\def \ch {\rm{ch}}

\def \cell {\rm{cell}}

\begin{document}

\title{Realistic Implementation of Chiral Magnetic Wave\\ in Heavy Ion Collisions}

\author{Ho-Ung~Yee}
\affiliation{
Physics Department,  University of Illinois,
Chicago, IL, 60607%
}
\affiliation{
RIKEN-BNL Research Center,  Brookhaven National Laboratory, Upton, New York 11973-5000}
\email{hyee@uic.edu}

\author{Yi~Yin}
\affiliation{
Physics Department,  University of Illinois,
Chicago, IL, 60607%
}
\email{yyin3@uic.edu}

\begin{abstract}

Chiral Magnetic Wave (CMW) is a gapless collective excitation of  chiral charges along the direction of magnetic field in the Quark-Gluon Plasma that 
arises from the triangle anomaly of QCD. We perform reliable study of the CMW in the realistic simulation of heavy ion collisions, and
 find that the CMW contributions to the charge dependent elliptic flow of pions, 
$\Delta v_2\equiv v_2(\pi^-)-v_2(\pi^+)$, 
linearly depending on the net charge asymmetry $A_\pm\equiv (N_+-N_-)/(N_++N_-)$ with a positive slope $r$, 
is comparable to the recent experimental results from RHIC. 
We identify ``Freezeout Hole Effect'', which is a direct consequence of the propagation of CMW during the realistic evolution of fireball, 
as the dominant physics effect responsible for a sizable contribution from the CMW to the slope parameter $r$, and
emphasize that a proper treatment of the freeze out condition is crucial in any reliable computation of the CMW contribution to the slope parameter $r$.
We also implement chiral phase transition effect in our study, which illustrates the sensitivity of the results to chiral phase transition temperature, 
and suggest that the CMW can be an important probe of QCD chiral phase transition. 
Our results on the impact parameter dependence compare well with the RHIC experiments. 
We also give predictions for the LHC energy.  

\end{abstract}

\preprint{RBRC-1050}

\maketitle

\section{Introduction}

Heavy Ion Collision experiments in RHIC and LHC are the unique opportunities to experimentally realize the new state of QCD matter with extreme high temperature, so that the fundamental
constituents of nuclear matter, quarks and gluons, are liberated from color confinement to form a quark-gluon plasma (QGP). Many experimental evidences collected from the on-going RHIC and LHC indicate a strongly coupled nature of the QGP, which has provided much motivation and many challenges in understanding the properties of the created QCD plasma.
Although the very early stage of the plasma at the time much
less than 1 fm is still beyond our complete theoretical control, hydrodynamics has been used to model the evolution
after the initial stage to successfully explain majority of experimental observations with a few transport coefficients without worrying too much about microscopic details of the QCD at strong coupling. 

Hydrodynamics is based on the conservation equations, and this is why triangle anomaly (chiral anomaly) of axial symmetry conservation affects some of hydrodynamic properties of charge fluctuations in QGP. 
Chiral Magnetic Effect (CME) \cite{Kharzeev:2007jp,Fukushima:2008xe,Son:2004tq}, which dictates a vector (axial) current along the magnetic field in the presence of axial (vector) charge due to triangle anomaly,  leads  to  an  important  modification of the hydrodynamic constitutive relations of the vector and axial charge currents \cite{Son:2009tf}:
 \be
 \label{eq:jmu}
J^\mu_{\rm{V},\rm{A}}=n_{\rm{V},\rm{A}} u^{\mu}
+\Delta J^{\mu}_{\rm{V},\rm{A}} +\ldots\, ,
\ee
where $\ldots$ refers to viscous corrections,
and $\Delta J^{\mu}_{\rm{V},\rm{A}}$ denotes the (axial) current induced by CME: 
\be
\label{eq:CME1}
\Delta J^{\mu}_V=\frac{eN_c}{2\pi^2}\mu_A B^{\mu}\, ,
\qquad
\Delta J^{\mu}_A=\frac{eN_c}{ 2\pi^2}\mu_V  B^{\mu}\, ,
\ee
where $B^\mu=(1/2)\epsilon^{\mu\nu\alpha\beta}u_\nu F_{\alpha\beta}$ is the magnetic field boosted from the fluid rest frame, and
$\mu_V,\mu_A$ are vector and axial charge chemical potentials respectively.
In heavy ion collisions, axial charges $\mu_A$ may be generated event-by-event through glasma color fields or sphaleron transitions, and the magnetic field created by the spectator charges of ultra relativistic heavy ions can be as large as $eB\sim 10 m_\pi^2$, and it points to the perpendicular direction of the reaction plane \cite{Kharzeev:2007jp}. 
These conditions make the CME possible
in heavy ion collisions, which would induce event-by-event charge separation perpendicular to the reaction plane. Experimental signatures in two particle correlations of $\cos(\phi_1+\phi_2)$ between same charged or opposite charged particles that are sensitive to this charge separation \cite{Voloshin:2004vk} are in favor of the prediction of the CME \cite{Abelev:2009ac,Selyuzhenkov:2011xq}, but there are other unexplained features in the experiments and the background effects unrelated to the CME \cite{Bzdak:2009fc,Wang:2009kd,Asakawa:2010bu,Pratt:2010zn}, so that the situation is not conclusive\cite{Bzdak:2012ia}.

Writing down the anomalous part of \eq\eqref{eq:CME1} in the absence of background flow,
\be
\label{eq:CME}
\vec J_V=\frac{eN_c}{2\pi^2}\mu_A \vec B\,,\quad \vec J_A=\frac{eN_c}{2\pi^2}\mu_V \vec B\,,
\ee
one obtains a new gapless hydrodynamic propagating mode of chiral charges, coined as Chiral Magnetic Wave (CMW) \cite{Kharzeev:2010gd,Newman:2005hd}.
A distinctive feature of the CMW is that it is about how charge fluctuations of arbitrary shape propagate along the direction of the magnetic field, and it does not require the presence of  background axial or vector charge densities. Defining left (right)-handed chiral charges by $J_{V,A}=\mp J_L+J_R$, these chiral charge fluctuations 
propagate according to the hydrodynamic dispersion relation \cite{Kharzeev:2010gd}
\be
\omega=\mp v_\chi k -i D k^2+\cdots\,,
\ee
where the momentum $k$ is along the direction of the magnetic field, and the Chiral Magnetic Wave velocity is determined in terms of the chiral charge susceptibility, $\chi_{L,R}$, of the plasma as\cite{Kharzeev:2010gd}
\be
v_\chi=\frac{eN_c B}{ 4\pi^2 \chi_{L,R}}
= \frac{eN_c B}{2\pi^2\chi_{V}}\, ,
\label{vchi}
\ee
with $\chi_{V}$ the vector charge susceptibility. 
Note that the sign in front of the first term is dictated by the chirality of the charge fluctuations: the CMW is uni-directional depending on the chirality of the charge fluctuations.
However, it should be stressed that the direction of motion does not depend on the sign of the fluctuations of a given chirality: any profile of fluctuations, either positive or negative, of a given chirality move to the same direction. In this sense, the CMW is not simply a separation of net axial charges $J_A=-J_L+J_R$.

\begin{figure}[t]
	\centering
	\includegraphics[width=0.32\textwidth]{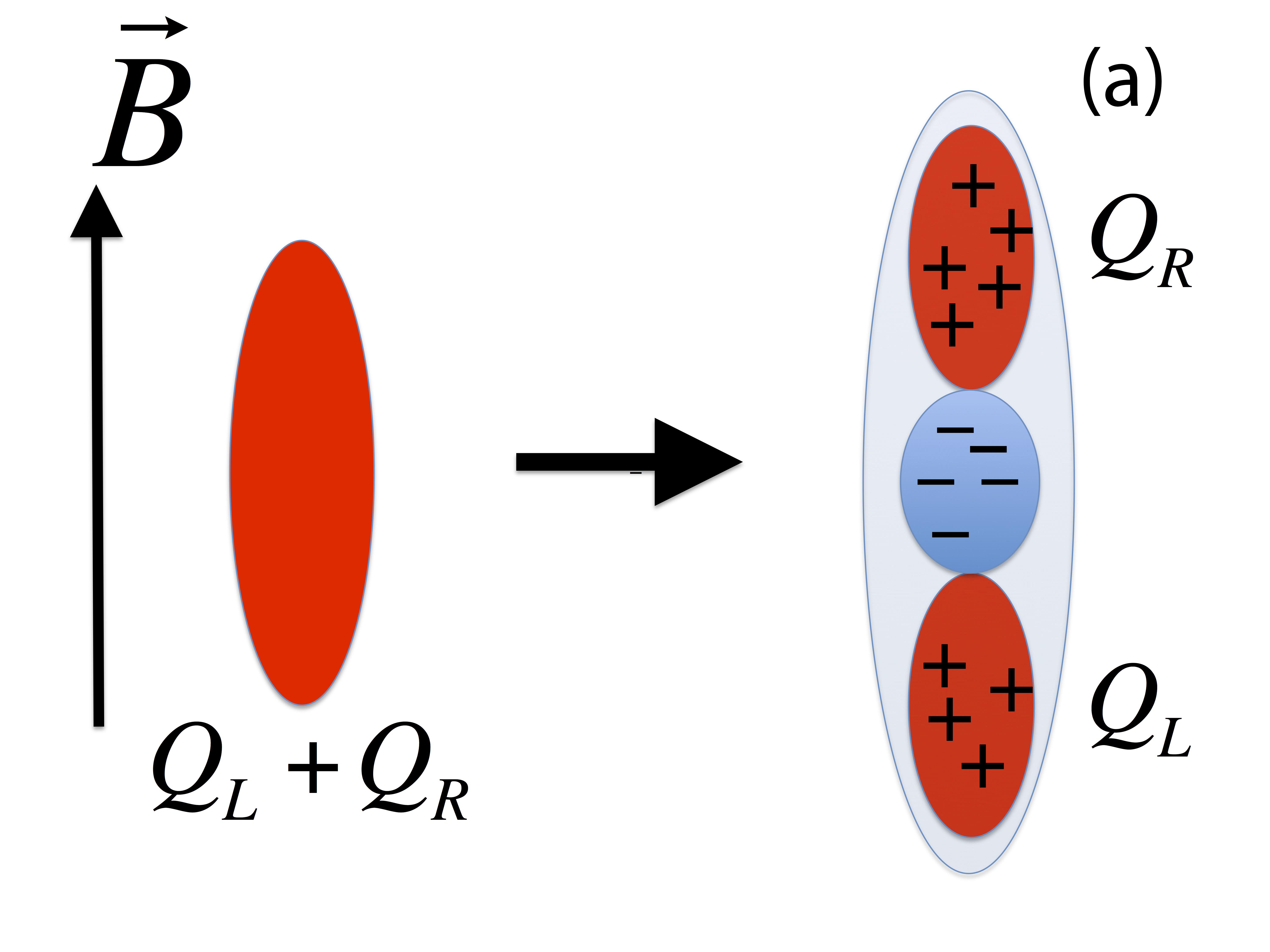}
        \includegraphics[width=0.65\textwidth]{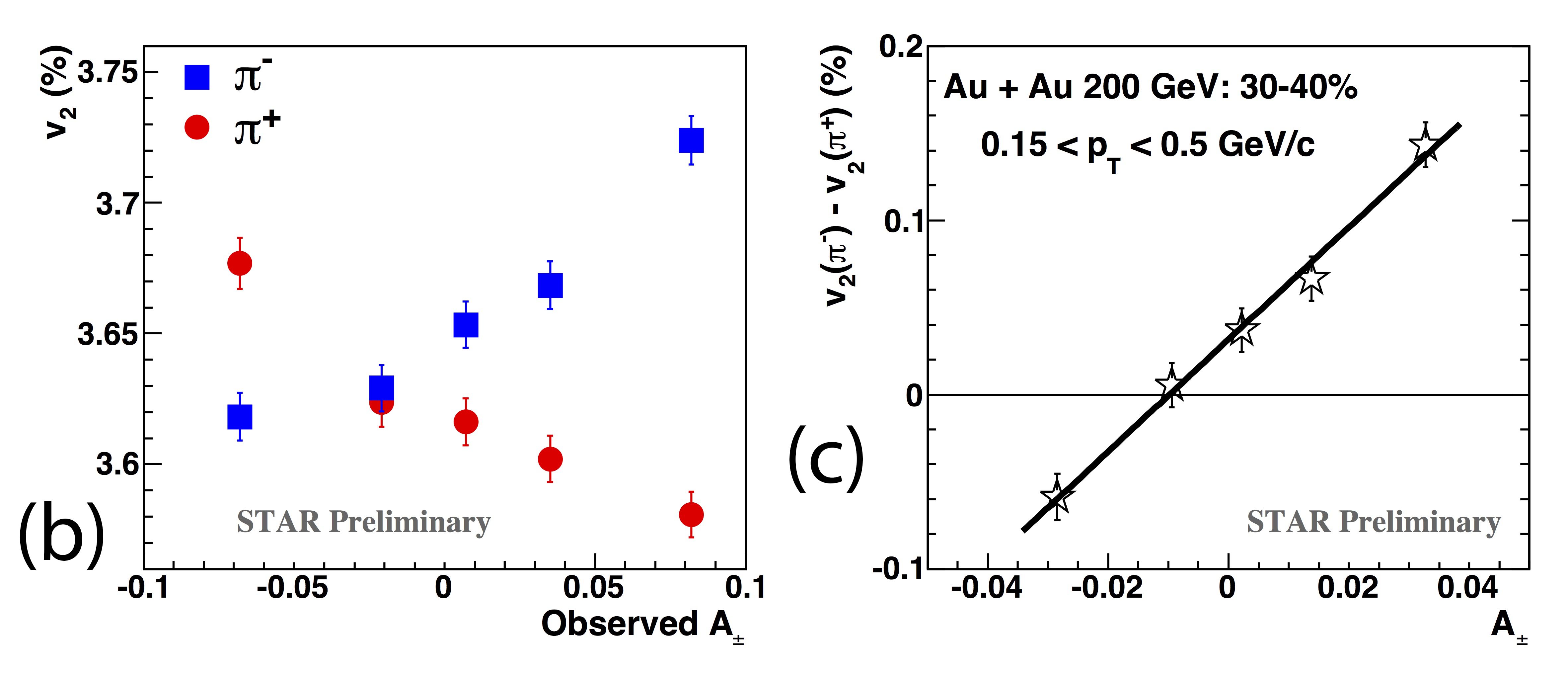}
		\caption{ (Color online) The Chiral Magnetic Wave induces an electric quadrupole moment proportional to the initial charge asymmetry ({\it left}), which eventually leads to the charge dependent elliptic flow of pions $\Delta v_2\equiv v_2(\pi^-)-v_2(\pi^+)$. 
The RHIC experiments\cite{Wang:2012qs,Ke:2012qb} confirmed the linear dependency of $\Delta v_2$ on the charge asymmetry $A\equiv (N_+-N_-)/(N_++N_-)$ ({\it middle and right}). The intercept at $A=0$ was explained in Ref.\cite{Stephanov:2013tga} using electric fields.
	\label{fig1}}
\end{figure}
It was suggested that one of the experimental signatures of the CMW in heavy ion collisions is the charge dependent elliptic flow of pions, $\Delta v_2\equiv v_2(\pi^-)- v_2(\pi^+)$, linearly
depending on the net charge asymmetry of the plasma $A_\pm\equiv (N_+-N_-)/(N_++N_-)$, that is, $\Delta v_2=r A_\pm$ with a positive slope  $r>0$ \cite{Burnier:2011bf,Burnier:2012ae}.
It is based on the fact that the net charge asymmetry (with zero axial charge in average) is a linear superposition of equal amount of left- and right-handed chiral charges via the relation $J_{V,A}=\mp J_L+J_R$. In the presence of magnetic field, these chiral charges move along the magnetic field in opposite directions to each other according to the CMW, resulting in an excess of (chiral) charges
around the pole regions of the plasma fireball and a depletion of charges in the central region, which leads to a net electric quadrupole moment proportional to the initial charge asymmetry $A_\pm$. See Figure \ref{fig1} ({\it left}). A similar observation was also made in Ref.\cite{Gorbar:2011ya}. This spatial distribution of charge asymmetry together with the background radial flow eventually leads to a difference between elliptic flows of positive and negative pions. 
The positive sign of the slope $r>0$ is an important prediction of the CMW. Recent experimental analysis of the RHIC data confirmed the linear dependence with a positive slope $r>0$ (see Figure \ref{fig1} ({\it middle and right})).

\begin{figure}[t]
	\centering
	\includegraphics[width=0.6\textwidth]{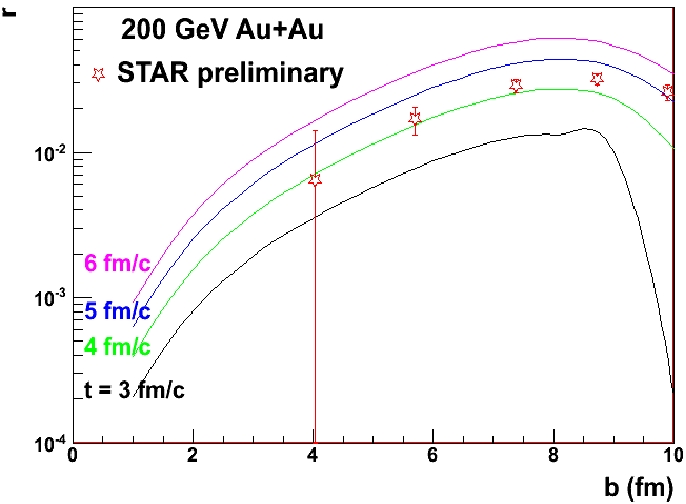}
		\caption{(Color online) The results of Refs.~\cite{Burnier:2011bf,Burnier:2012ae} for the impact parameter dependence of slope parameter $r$ with different lifetimes of the magnetic field (different colored lines). 	The stared points are from the RHIC experiments \cite{Wang:2012qs,Ke:2012qb}. \label{fig2}}
\end{figure}

In Refs.~\cite{Burnier:2011bf,Burnier:2012ae},
 numerical simulations of the CMW were performed with a few simplified conditions for an estimate purpose:
\begin{enumerate}
\item It was a 2+1 dimensional simulation with the static plasma and the magnetic field lasting for a finite time before applying blast wave approximation;

\item The transverse profile of the temperature obtained from the KLN model with the Wood-Saxon nuclear shape was used;

\item Initial charge asymmetry was distributed homogeneously in the transverse plane;
 
\item No realistic freeze out condition was used;

\item The CMW velocity from the strong coupling AdS/CFT model was used inside the boundary where the temperature crosses the chiral phase transition. On the boundary, the CMW velocity was put to zero since the CMW disappears in the chiral symmetry broken phase .
 \end{enumerate}
Despite these simplifications, the numerical results in Refs.~\cite{Burnier:2011bf,Burnier:2012ae} were able to explain the experimental results at RHIC with a tuning of the strength and the lifetime of the magnetic field.  Somewhat surprisingly, the simulation also explained the impact parameter dependence of the slope $r$ quite well, which is not trivial. See Figure \ref{fig2}.

In this work, we perform more realistic simulation of the CMW improving all five points mentioned in the above.
 We also point out several important physics aspects in the simulation that 
have to be taken into account, in order to get a trustable numerical result for the CMW contribution to the slope parameter $r$.
Especially, we identify 
``Freezeout Hole Effect'' happening in the realistic simulation as the dominant source of  the CMW contribution to the slope $r$.
In heavy-ion collisions,
the (chiral) charge flow of a given fluid cell is heavily deflected to the direction of the magnetic field due to the CMW. 
The initial charges of a given chirality will move to a definite direction along the magnetic field before hitting the freeze out surface, 
so that a part of the freeze out surface that lies in front of the motion induced by the CMW would encounter more charges hitting the surface, 
while the freeze out surface in the opposite side would see much less charges, and there exists a part of the freeze out surface that sees no charges of that given chirality at all: we call it ``Freezeout Hole''. 
Adding contributions from both chiralities would result in an ``quadrupole-like" charge distribution on the freeze-out surface.
We find that this ``Freeze-out Hole" Effect is the dominant mechanism of the CMW contribution to the slope $r$.

Another important ingredient we implement is the chiral phase transition effect.
Near the region of the chiral phase transition, the CMW velocity should naturally drop to zero, since the CMW disappears in the chiral symmetry broken phase. 
Note that the chiral phase transition region is inside the chemical freeze out surface. The charge flow of a given fluid cell will be dramatically different before and after hitting the chiral phase transition region. 
This jump of the CMW velocity across the chiral boundary
significantly modifies the charge flow of fluid elements and the final charge distribution on the freeze out surface.


Our result points to that the slope parameter $r$ receives a sizable contribution from the CMW which is comparable to (though somewhat less than) the experimental value.
Note that the slope $r$ is a P-even observable that is subject to other possible effects unrelated to triangle anomaly.
Our findings are in contrast to those of the recent simulation in Ref.~\cite{Hongo:2013cqa} which claims that the contribution to $r$ from the CMW is negligibly small.
One possible reason for this discrepancy  seems to be a crude treatment of 
the freeze out condition
in Ref.~\cite{Hongo:2013cqa}:
the freeze out surface is taken at a fixed time. The chiral phase transition effect to the CMW propagation seems to be absent in 
Ref.~\cite{Hongo:2013cqa} too.
The work of Ref.\cite{Hongo:2013cqa} was certainly a very important step towards a realistic numerical study of the anomalous hydrodynamics, but we inevitably disagree on a few important details, since the slope $r$ is very sensitive to them. With our initial condition which seems well justified as we explain in the main text, we don't find a large contribution to the slope $r$ without the CMW. 

Our paper is organized as follows. 
In section \ref{sec2}, 
we describe first the CMW in a non-static inhomogeneous expanding plasma in an ideal hydrodynamic limit, neglecting viscosities and charge diffusions. 
We develop a new and more intuitive way of describing the charge flow dynamics in an ideal limit, 
based on the concept of co-moving fluid cells, 
where the local charge conservation is manifestly realized. 
The result obtained in this way can also be interpreted as the Green's function of charge density connecting the initial surface to the final freeze out surface, 
that can be easily integrated over any initial conditions,
 so it should be very useful by itself for future studies. 
We then explain in detail the various elements of our realistic numerical simulation of the CMW. 
In section \ref{sec3}, we highlight the
``Freezeout Hole Effect'' and 
discuss the role played by the chiral phase transition (cross-over),
by providing our numerical evidences for their importance. 
In section \ref{sec4}, we present our numerical results, 
including the dependence on the impact parameter, 
the magnetic field and its lifetime.
We point out that the ratio $\kappa = r/\bar{v}_2$, where 
$\bar{v}_2 =(v^{+}_2+v^{-}_2)/2$ is the background (charge-independent) elliptic flow, is a convenient experimental observable characterizing the CMW contribution to the charge-dependent elliptic flows. 
We also predict that at LHC energy the dependence of $\kappa$ on impact parameter $b$ would be different from that in RHIC and offer an explanation for that difference.

While our work was near the final stage, 
Ref.~\cite{Taghavi:2013ena} appeared which studied the CMW in an 
Bjorken-like expanding plasma.
Our formalism of CMW in a general hydrodynamic background is consistent with Ref.~\cite{Taghavi:2013ena}.
Ref.~\cite{Taghavi:2013ena} discussed the smallness of the magnitude of the snapshot of the charges due to the CMW, 
while the slope $r$ of our interest is sensitive to the
integrated charge flow across the freeze out surface.
We think this is one reason that some of our conclusions seem different from those in Ref.~\cite{Taghavi:2013ena}.

\section{Realistic Implementation of Chiral Magnetic Wave\label{sec2}}

We first describe the formulation of the CMW charge flow in a general ideal hydrodynamic background where the charge density is treated as a small linearized perturbation to the background plasma. Instead of solving the linearized hydrodynamic equation for the charge perturbations directly (we will write this down any way shortly), we find an alternative description in terms of co-moving fluid cell of charges more intuitive and useful in our numerical study.
We mention that the usefulness of this description is limited to the ideal case without diffusion.

Let us illustrate the idea in the case of no magnetic field or CMW. For an ideal case, the constitutive relation of charge current reads as
\be
J^\mu=n u^\mu\,,
\ee
with the conservation equation 
\be
\label{eq:nconservation}
\partial_\mu J^\mu=\partial_\mu\left(n u^\mu\right)=\left(u^\mu\partial_\mu\right)n+n\left(\partial\cdot u\right)=0\,.
\ee
The first term is the co-moving variation of the charge density while the second term represents the expansion of the volume of the co-moving fluid cell.
The above equation means that the total charge inside the walls of a co-moving fluid cell does not change: it is the charge conservation without any diffusion across the walls.
This implies that any points along the trajectory of a given fluid cell 
given by solving the first order equation
\be
\frac{d X^\mu(\lambda)}{d\lambda}=u^\mu(X)\,,\label{backtraj}
\ee
has the same total charge inside the fluid cell.
Each trajectory can be interpreted as a mapping from the initial surface to the final freezeout surface,
and therefore, these trajectories can also be thought of as defining a Greens function transporting initial data to the final freeze out surface.

We next turn to the case of our interests: the charge flow with the CMW,
in the above picture of the co-moving fluid cell. 
Because the charge flow is affected by the CMW velocity, the co-moving velocity that traces the chiral charge flow deviates from the background fluid velocity $u^\mu$. To see this more clearly, let's start from the constitutive relation of chiral charge currents defined by
$J_{L,R}^\mu=(J_V^\mu\mp J_A^\mu)/2$,
\be
J_{L,R}^\mu=n_{L,R}u^\mu\mp \frac{eN_c \mu_{L,R}}{ 4\pi^2} B^\mu+\cdots\,,
\ee
where, as a reminder, $B^\mu=(1/2)\epsilon^{\mu\nu\alpha\beta}u_\nu F_{\alpha\beta}$ is the magnetic field boosted from the fluid rest frame, and the second term is the CME contribution.
Since we are treating the charges as linearized perturbations, the chemical potentials at the linearized level are related to the density fluctuations $n_{L,R}$ by
\be
n_{L,R}=\chi_{L,R}(T) \mu_{L,R}
= \frac{1}{2}\chi_{V}\mu_{L,R}
\, ,
\ee
with the susceptibility $\chi_{L,R}(T)$ which are equal for both chiralities as the QCD is P-conserving. 
Defining the CMW velocity in a general boosted frame,
\be
u^\mu_{\chi}\equiv \frac{eN_c}{ 4\pi^2 \chi_{L,R}(T)}B^\mu
= \frac{eN_c}{2\pi^2\chi_{V}}B^{\mu}\, ,
\ee
the linearized constitutive relations become
\be
J^\mu_{L,R}=n_{L,R}\left(u^\mu\mp u^\mu_{\chi}\right)\equiv\tilde n_{L,R}\tilde u_{L,R}^\mu\,,
\ee
where
\be
\tilde u_{L,R}^\mu\equiv \frac{(u^\mu\mp u^\mu_{\chi})}{\sqrt{1-v_\chi^2}}\,,
\ee
is the properly normalized 4-vector $g_{\mu\nu}\tilde u^\mu_{L,R} \tilde u^{\nu}_{L,R}=-1$, and $\tilde n_{L,R}=n_{L,R}\sqrt{1-v_\chi^2}$ is the density in the rest frame of $\tilde u_{L,R}^\mu$, and
\be
v_\chi=\frac{eN_c}{ 4\pi^2 \chi_{L,R}(T)}B
= \frac{eN_c}{ 2\pi^2 \chi_{V}(T)}B\, ,
\ee
is the CMW velocity in the local rest frame\cite{Kharzeev:2010gd}. 
We used $u^\mu_{\chi} u_\mu=0$ and $g_{\mu\nu}u^\mu_{\chi} u^{\nu}_{\chi}=v_\chi^2$. 
The new velocity fields $\tilde u^\mu_{L,R}$
for left or right-handed chiralities are the effective velocity fields for the chiral charge motion affected by the CMW contribution $u^\mu_{\chi}$.
The conservation equations then give
\be
\partial_\mu J^\mu_{L,R}=\partial_\mu\left(\tilde n_{L,R} \tilde u^\mu_{L,R}\right)=0\,,
\ee
which simply means that we can apply the previous co-moving fluid cell picture of charge conservation with the modified fluid velocity $\tilde u^\mu_{L,R}$ and the chiral charge densities $\tilde n_{L,R}$ for each left-handed (L) or right-handed (R) chirality separately.
This generalizes the CMW to an arbitrary hydrodynamic background.
Therefore, the trajectories, determined by solving the first order equation
\be
\frac{d X^\mu(\lambda)}{ d\lambda}=\tilde u^\mu_{L,R}(X)\,,\label{CMWtraj}
\ee
with $\lambda$ being the affine parameter of the trajectories,
can be thought of as describing the motion of chiral charges in the background
plasma. 
The trajectories from (\ref{CMWtraj})
give a mapping from an initial surface to the final freeze out surface, 
transporting the initial data to the final chiral charge densities, 
which defines a Green's function of the chiral charge densities in the presence of the CMW. 
The fluid cell picture co-moving with $\tilde u_{L,R}^\mu$ is an intuitive way of looking at the charge motion affected by the CMW.

The above fluid cell picture provides us with a convenient numerical method to solve the charge flow dynamics.
We assume boost invariance and the initial condition at the Bjorken time $\tau=\tau_i$ is specified by the charge density profile on the transverse 2-plane ($x,y$).
We divide the initial transverse space into small squares of equal area, 
say $A^{i}_{cell}$,
they define our fluid cells and we numerically 
integrate 
(\ref{CMWtraj}) to find the stream-line trajectories of each fluid cells.
This gives us the chiral charge motion with the CMW and the volume expansion of each fluid cells. 
Note that the {(chiral)}
charge conservation ensures that the total (chiral) charge inside each cell is constant, 
given by the initial charge distribution. 
Using this fact and the volume expansion, 
one can find the charge density at later times.
When the charge cell trajectory hits the freeze-out surface (of constant temperature $T=T_f$ in our analysis), 
we obtain the charge dependent particle spectrum (more precisely, the difference between the positive and negative pion momentum spectrum) emitted from the cell using the Cooper-Frye type treatment (more detail in the following), 
and we sum over all the cells to find the net difference in the positive and negative pion momentum distributions.

We now 
 explain our freeze out treatment in more detail.
We first consider a fluid cell which is initially located at $(x_i, y_i)$ at $\tau=\tau_i$
(say for example, point E in Fig.~\ref{fig:streamline}).
We denote its initial vector charge density by $n^{i}_{\cell}$.
In the present work,
we consider the case where there is no axial charge initially, so that  $n^{i}_{L}=n^{i}_{R}= n^{i}_{\cell}/2$.
We then track the trajectories of left-handed (right-handed) charges determined by \eq\eqref{CMWtraj}.
When the trajectory hits the freeze-out surface, 
we denote the corresponding fluid velocity as $u^{\mu}_{f, L}$ ($u^{\mu}_{f,R})$ (c.f.~Fig.~\ref{fig:streamline}). 
The momentum distribution of pions 
due to the fluid cell we are considering
is given by the boosted thermal distribution summed over both left-handed and right-handed contributions,
\be
p^0\frac{d^3 N^\pm_{\cell}}{d^3  p}=\frac{d^3N^\pm_{\cell}}{ p_\perp dp_\perp dY d\phi}=
\frac{p_\mu }{(2\pi)^3} 
\[\, V^f_{\rm cell,L}\,u^\mu_{f,L}\, e^{-{(p_{\mu} u^{\mu}_{f,L}\mp \mu^{f}_{L})\over T_f}}
+ V^f_{\rm cell,R}\,u^\mu_{f,R}\, e^{-{(p_{\mu} u^{\mu}_{f,R}\mp \mu^{f}_{R})\over T_f}}
\],\label{N}
\ee
where $p_\perp=\sqrt{p_x^2+p_y^2}$, 
and $\phi$ is the azimuthal angle in $(p_x,p_y)$ plane, and 
\be
Y={1\over 2}\log\left (p^0+p^z\over p^0-p^z\right)\,,
\ee
is the momentum rapidity. 
The $V^f_{\rm cell, L}$($V^f_{\rm cell, R}$) is the 3-volume of the cell in its local rest frame. 
For $p_\perp\gg m_{\pi}$, 
one can safely replace the Bose-Einstein distribution with the Boltzmann distribution.

A clarification on the notations of $n^{f}_{L}, n^{f}_{R}$ or $\mu^{f}_{L}, \mu^{f}_{R}$ is due here.
Note that the meaning of chirality (left or right-handed) disappears 
beyond the chiral phase transition surface since the chiral symmetry is heavily broken by the condensate. 
What remains meaningful is the net vector (electric) charge which is a simple sum of left and right-handed chiral charges.
Beyond the chiral phase boundary, the charge flow is the usual ideal charge flow without CMW which is the same, independent of chirality,
i.e. $\tilde{u}^{\mu}_{L,R}=u^{\mu}$. 
Since we are working on the linearized charge fluctuations, 
the superposition principle applies, 
and we can treat the vector charges originating from the left or right-handed chiral charges separately: 
the final total electric charge on the freeze out surface would be a simple sum of them. 
Therefore,
when we say the left-handed(right-handed) charge (chemical potential) beyond the chiral phase boundary, 
what we really mean is the vector charge(chemical potential) originating from the left-handed(right-handed) charge inside the chiral transition boundary and flowing according to \eq\eqref{CMWtraj}.

One may define the charge dependent distribution of a fluid cell, $N^{\ch}_{\cell}$, by $N^{\ch}_{\cell}=(N^{+}_{\cell}-N^{-}_{\cell})/2$.
We then have, 
to linear order in $\mu_f/T_f$,
\begin{multline}
p^0{d^3 N^{\ch}_{\cell}\over d^3  p}
=
\frac{1}{T_f}{p_\mu \over (2\pi)^3} 
\[\, \mu^{f}_{L}\, V^f_{\rm cell,L}\,u^{\mu}_{f,L} e^{-{p_{\mu} u^{\mu}_{f,L}\over T_f}}
+ \mu^{f}_{R}\, V^f_{\cell,R}\,u^\mu_{f,R} e^{-{p_{\mu} u^{\mu}_{f,R}\over T_f}}\, \]=
\\
\frac{2}{\chi_{V,f} T_f}
\frac{p_\mu }{ (2\pi)^3} 
\[\, n^{f}_{L}\,V^f_{\cell,L}\,u^{\mu}_{f,L} e^{-{p_{\mu} u^{\mu}_{f,L}\over T_f}}
+ n^{f}_{R}\, V^f_{\cell,R}\,u^{\mu}_{f,R} e^{-{p_{\mu} u^{\mu}_{f,R}\over T_f}}\, 
\]\,,
\end{multline}
where $\chi_{V,f} = \chi_{V}(T_f)$ is the susceptibility evaluated at freeze-out temperature.
Because the total (chiral) charge inside a cell is conserved along the streamline, 
we have
\be
V^f_{\rm cell,L} n^{f}_{L}=
\tilde V^i_{\rm cell, L} \tilde n^{i}_{L}= \frac{1}{2}V^i_{\rm cell}n^{i}\, ,
\qquad
V^f_{\rm cell,R} n^{f}_{R}
= \frac{1}{2}V^i_{\rm cell}n^{i}\,,
\ee
where $\tilde V^i_{\rm cell,L} (V^i_{\rm cell})$ is the volume in the rest frame of $\tilde u^\mu_{L} (u^\mu)$ which is given by the initial data, and moreover the initial 3-volume, which is at rest initially along the transverse directions, is simply
\be
V^i_{\rm cell}=A^i_{\rm cell}\tau_i \Delta \xi\,,
\ee
where $\xi=\tanh^{-1}(z/t)$ is the spatial rapidity and $A^i_{\rm cell}$ is the initial transverse area of a square cell. Then we have for each cell
\be
\label{eq:Nchcell}
p^0{d^3 N^{\rm ch}_{\cell}\over d^3  p}=
\frac{1}{ (2\pi)^3}
\frac{A^i_{\rm cell}\tau_i n_i}{ T_f\chi_{V,f}} \int d\xi \, 
 p_\mu \, \(\,u^\mu_{f,L} e^{-{p_{\mu} u^{\mu}_{f,L}\over T_f}}
 +u^\mu_{f,R} e^{-{p_{\mu} u^{\mu}_{f,R}\over T_f}}\, \)\, ,
\ee
where we have integrated over the space rapidity for each transverse initial position of a cell. The $\xi$-integration can be done analytically. By writing
$u^\mu=(u^0,u^z,u^x,u^y)=(u^\tau\cosh\xi,u^\tau\sinh\xi,u^r\cos\phi_u,u^r\sin\phi_u)$
and
$p^\mu=(m_\perp\cosh Y,m_\perp\sinh Y,p_\perp\cos\phi,p_\perp\sin\phi)$
with $m_\perp=\sqrt{p_\perp^2+m_\pi^2}$,
the $\xi$-integration gives
\bear
\int d\xi \, p_\mu u^\mu e^{-{p_\mu u^\mu\over T_f}}=e^{{p_\perp\over T_f}u^r\cos(\phi-\phi_u)}
\left[
u^{\tau} m_\perp K_1\left({m_\perp u^\tau\over T_f}\right)-u^{r}p_\perp\cos(\phi-\phi_u) K_0\left({m_\perp u^\tau\over T_f}\right)\right]\,,\nonumber\\
\eear
which considerably 
helps to reduce the computation time of our analysis.
What remains is the sum over all transverse area elements:
\begin{equation}
p^0{d^3 N^{\rm ch}\over d^3  p}
=\sum_{\cell}\, p^0{d^3 N^{\rm ch}_{\cell}\over d^3  p}\, .
\end{equation}
What is nice about the above formula is that it is computed via a sum over ``initial'' surface area.

Mathematically, our freeze out treatment corresponds to taking $V^f_{\rm cell}u^\mu$ as the normal surface vector in the usual Cooper-Frye formula. 
In other words, the 3-dimensional ``freeze out surface'' is the 3-volume $V^f_{\rm cell}$ in the local rest frame and the normal vector is $u^\mu$.
Although our treatment is not strictly equal to the original Cooper-Frye, the difference is checked to be numerically less than 10$\%$, and the uncertainly is smaller than those involved in the Cooper-Frye itself.

From $N^{\rm ch}$, it is straightforward to obtain the charge dependent elliptic flow of pions, 
$\Delta v_2=v_2(\pi^-)-v_2(\pi^+)$, 
of our interest as follows.
From our definition of $N^{\rm ch}$,
i.e., $N^{\pm}=\bar{N}\pm N^{\ch}$, 
we have
\be
p^0{d^{3} N^\pm\over d^3 p}={d^{3} N^\pm\over p_\perp d p_\perp dY d\phi}={d^3 \bar N\over p_\perp d p_\perp dY d\phi}\pm {d^3 N^{\rm ch}\over p_\perp d p_\perp dY d\phi}\,.
\ee
Harmonically 
expanding each term in azimuthal angle $\phi$ from the reaction plane,
\bear
{d^3 \bar N\over p_\perp d p_\perp dY d\phi}&=&\bar v_0\left(1+2\bar v_2\cos(2\phi)+\cdots\right)\,,\nonumber\\
{d ^3N^{\rm ch}\over p_\perp d p_\perp dY d\phi}&=& v_0^{\rm ch}\left(1+2 v_2^{\rm ch}\cos(2\phi)+\cdots \right)\,,\label{v2}
\eear
one can easily derive that the charge asymmetry $A_\pm=(N^+-N^-)/(N^++N^-)$ is given by $A=v_0^{\rm ch}/\bar v_0$, and the elliptic flows of charged pions are
\be
v_2^\pm={\bar v_0\bar v_2\pm v_0^{\rm ch}v_2^{\rm ch}\over \bar v_0\pm v_0^{\rm ch}}={\bar v_2\pm A v_2^{\rm ch}\over 1\pm A} = \bar v_2\mp  (\bar v_2-v_2^{\rm ch}) A+{\cal O}(A^2)\,,
\ee
up to linear in $A$.
Then, we have
\be
\Delta v_2=v_2^--v_2^+=2(\bar v_2-v_2^{\rm ch})A\equiv r\,A\,,
\ee with the slope parameter
\be
r=2(\bar v_2-v_2^{\rm ch})\,.
\ee
Therefore, the slope parameter is essentially the (twice of) difference between the average and charge dependent elliptic flows.
In addition,
as will be explained in detail in the upcoming section,
it would be convenient to introduce a new parameter $\kappa$ defined as
\begin{equation}
\label{eq:kappa}
\kappa \equiv \frac{2(\bar v_2-v_2^{\rm ch})}{\bar{v}_{2}}
= \frac{2(v^{-}_{2}-v^{+}_{2})}{v^{+}_{2}+v^{-}_{2}}
= \frac{r}{\bar{v}_{2}}\, ,
\end{equation}
to characterize the charge asymmetry in the elliptic flows.

\begin{figure}
  \centering
	\includegraphics[width=.45\textwidth]{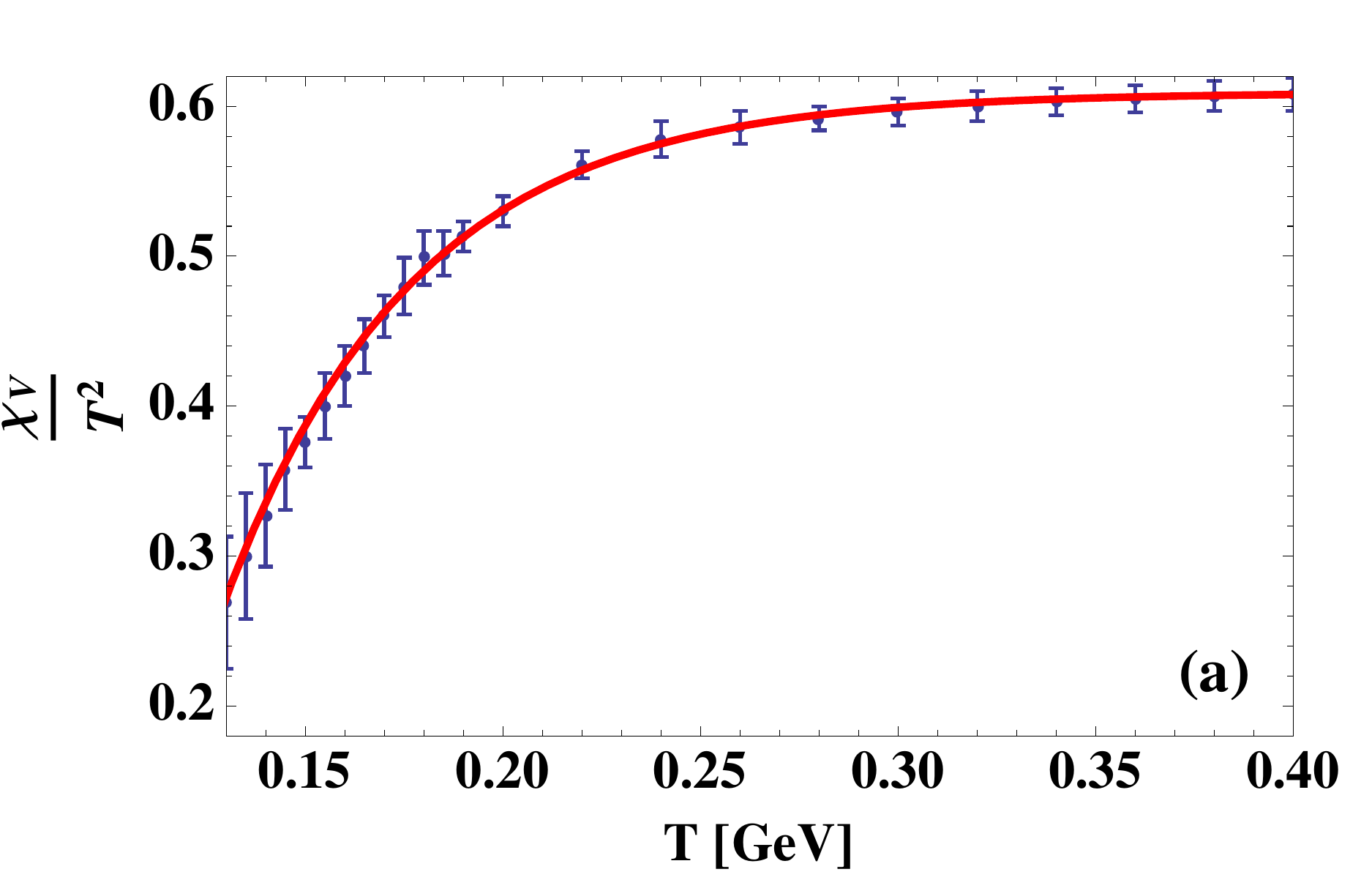}
	\includegraphics[width=.45\textwidth]{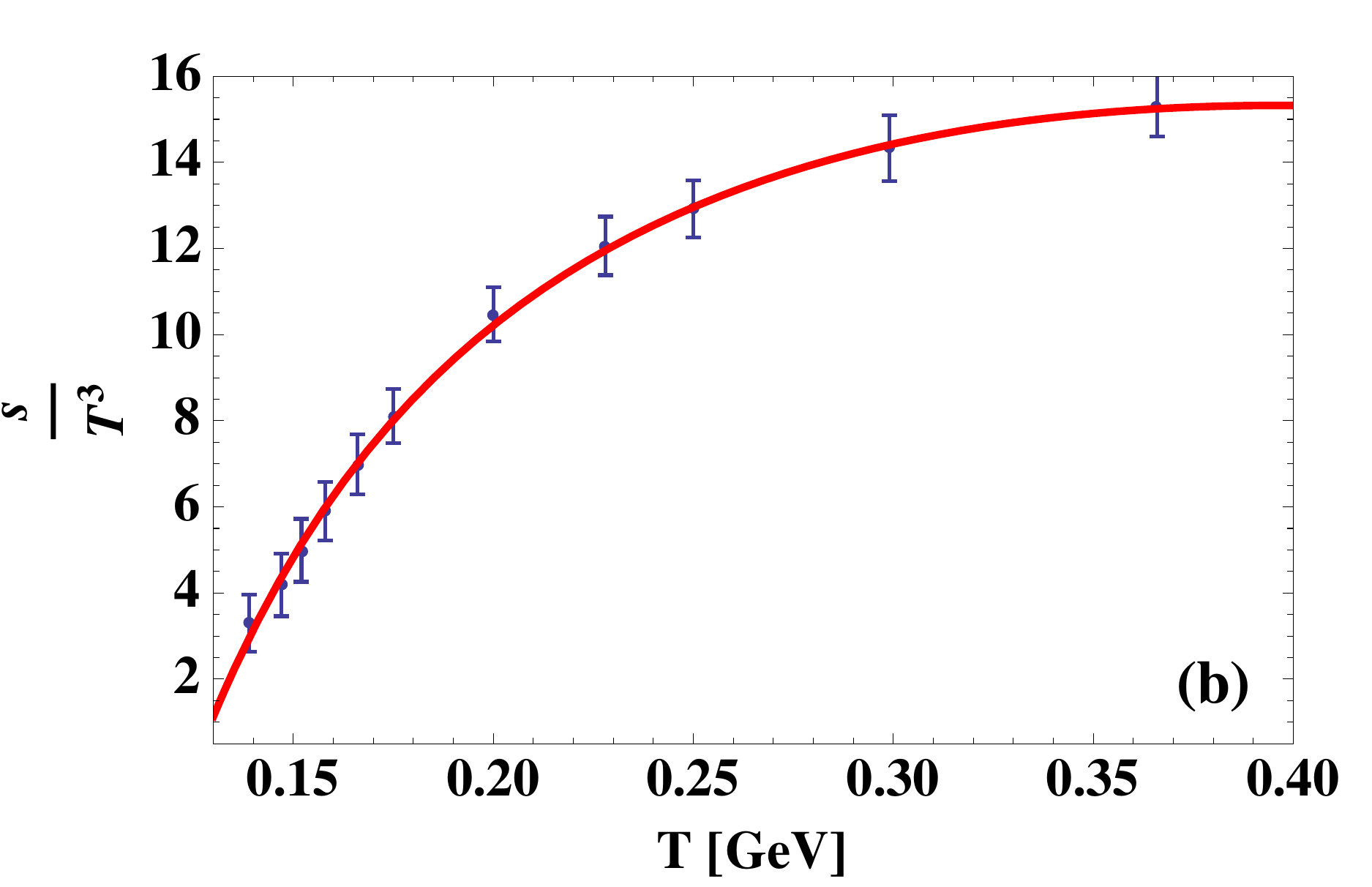}
	\caption{
          \label{fig:eos_fit}
(Color online) $\chi_{V}/T^{2}$(\textit{left}) and $s/T^{3}$(\textit{right}) measured in lattice\cite{Borsanyi:2010cj,Borsanyi:2011sw}.
In practice, 
we parametrize lattice data as $\chi_{V}/T^2 = 0.609 (1 - 8.29 e^{-\frac{T}{0.048}})$ and
$s/T^3 = (-17.4 - 0.185/T^{2}+62.7 T^{0.33}-46.2 T^{1.43})$ where in those expression, temperature $T$ is in GeV. 
Lattice data points with error bar are in blue and the results of above parametrization are shown in red solid curves.
        }	
 \end{figure}

We now detail the elements of our numerical analysis which improves the previous simulation in Refs.~\cite{Burnier:2011bf,Burnier:2012ae}  regarding the five points mentioned in the introduction.

\begin{enumerate}
\item We use the existing boost-invariant 2+1 dimensional (ideal) hydrodynamic simulation code by U.~Romatschke and P.~Romatschke \cite{Romatschke:2007mq,Luzum:2008cw} 
to generate a realistic background plasma evolution with the Glauber initial condition, on top of which the charge asymmetry of our interest
is treated as a linear perturbation, obeying the equation of motion derived in the previous section
\footnote{
We are grateful to P. \& U. Romatschke who made their codes accessible to the public. 
These codes are maintained by M. Luzum and can be downloaded via
\href{http://matt.luzum.org/hydro/}{http://matt.luzum.org/hydro/} .
}.
Changing the Glauber initial condition to the KLN (CGC) model does not change our results much. 
The initial time is set at $\tau_i=0.4$ fm.
It is worth pointing out that while the charge-independent $\bar{v}_2$ is not very sensitive to $\tau_i$ as the plasma has enough time to generate the elliptic flow \cite{Shen:2010uy},
a relatively later initial time would underestimate the contribution of the CMW to the
charge dependent $v_2$, since the magnetic field is strongest at early time. 
For the QCD equation of state, we use the current lattice result from the Wuppertal Collaboration
\cite{Borsanyi:2010cj}( See Figure.~\ref{fig:eos_fit}).

\item  We 
take $\vB$ in the lab frame along the $y$ direction and
use the time-varying profile of the magnetic field with a parametrization
\be
eB(\tau)={(eB)_{\rm max}\over 1+(\tau/\tau_B)^2}\,,
\ee where we call $\tau_B$ the lifetime of the magnetic field. This form fits well to the exact result neglecting plasma matter effects, and has been used in previous literature widely (see, for example, Ref.~\cite{Basar:2012bp}).
Our magnetic field is however still homogeneous in space, which is not significantly different from the exact solution in the fireball region.
In Bjroken's coordinates, $B^{\mu}$ explicitly reads as
\begin{equation}
\label{eq:bmu}
eB^{\tau} = eB u_{y}\cosh(\xi)\, ,
\qquad
eB^{\xi } = -\frac{eB}{\tau} u_{y} \sinh(\xi)\, ,
\qquad
eB^{y} = eB u_{\tau}\cosh(\xi)\, .
\end{equation}
As we are interested in the result at mid-rapidity,
we set $\xi=0$ in the expression of $B^{\mu}$ in practical calculations (see Ref.~\cite{Taghavi:2013ena} for a similar treatment).

\item For the initial charge distribution on the initial surface, 
we take a constant homogeneous profile of the dimensionless ratio $n/s$, where $n,s$ are charge and entropy density in the local rest frame.
For high initial temperature where the QCD is approximately conformal, this is almost equivalent to a homogeneous $\mu/T$ (which in fact agrees with the initial condition used in Ref.~\cite{Hongo:2013cqa}).
This is based on the expectation that both charge density and the entropy density at initial time should be governed by the same scale, 
saturation scale $Q_s$, 
so that the dimensionless ratio should be approximately constant over the transverse space. Note that $Q_s$ (and $n,s$) itself is a non-trivial function in the transverse space.
As explained in the next section, 
with this initial condition
the slope parameter $r$ would be zero in the absence of CMW. 
This makes a clean separation between contributions to $r$ due to an initial profile and that due to the propagation of CMW during the evolution of the fireball.

\begin{figure}[t]
	\centering
	\includegraphics[width=7cm]{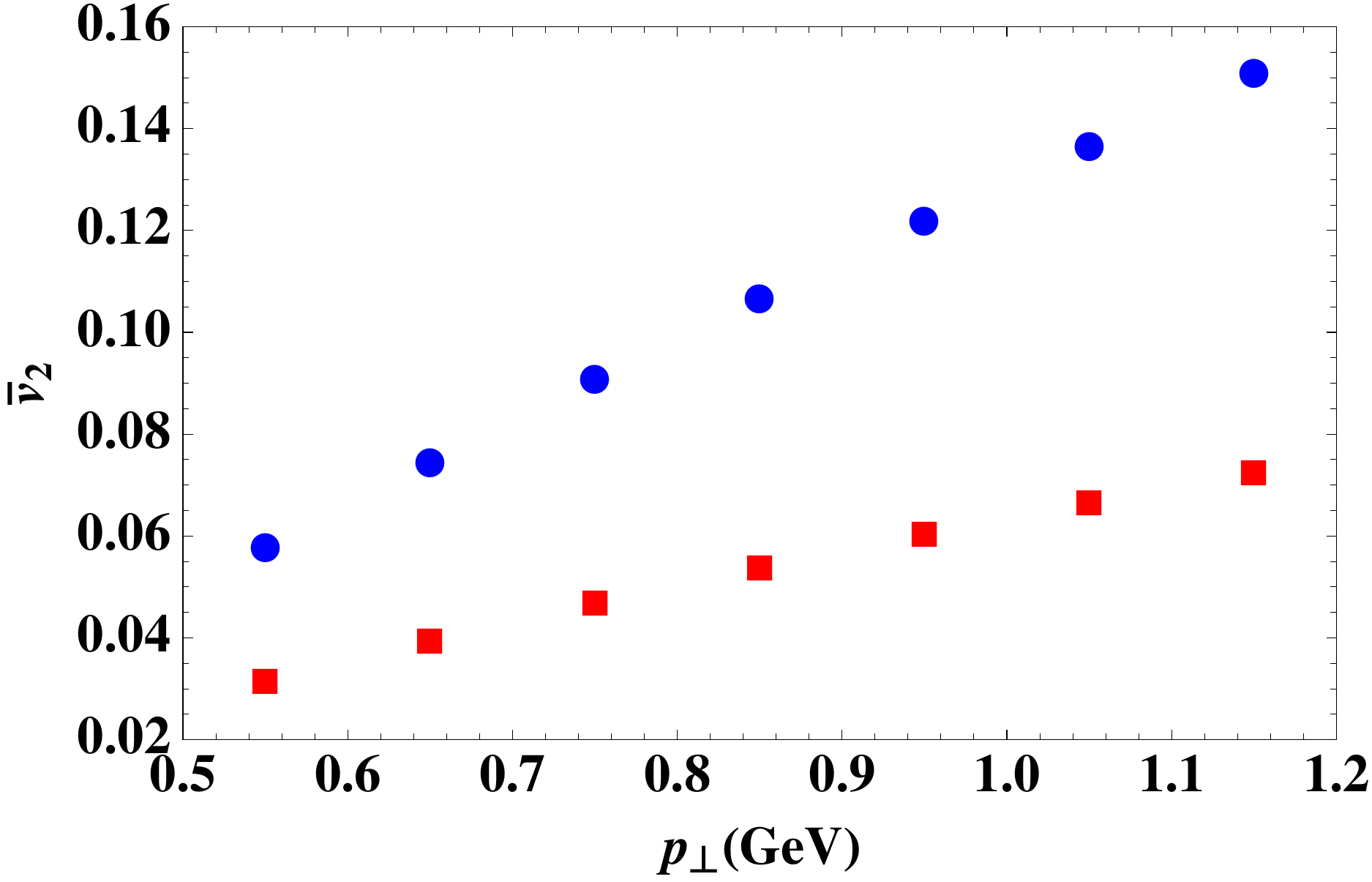}
		\caption{ (Color online ) Background pion elliptic flow $v_2$ as a function of transverse momentum in GeV (the upper dots for $b=7$ fm, the lower for $b=4$~fm). 	\label{fig3}}
\end{figure}
\item We use the Cooper-Frye type formula explained in the above with the constant temperature freeze out surfaces at $T=120$ MeV (guided by Ref.~\cite{Shen:2011eg}) to get the final momentum distribution of charged pions. 
The realistic plasma background we use reproduces the experimentally measured average pion elliptic flows as a function of transverse momentum well. This is shown in Figure \ref{fig3} for the RHIC energy $\sqrt{s_{NN}}=200$ GeV with impact parameters $b=4$ and $7$ fm, which are consistent with the experimental measurements at $0-20\%$ and $20-30\%$ centrality bins respectively\cite{Adams:2004bi}. This is an important prerequisite 
for our numerical results of the charge dependent pion elliptic flows to be trustable. We also implement the chiral phase transition effect by assuming a sharp chiral transition at $T=T_{\rm \chi}$
where the CMW velocity drops to zero discontinuously. Our results are somewhat sensitive to $T_{\chi}$ ranging from 150 MeV to 165 MeV, reducing our results up to factor 5 for $T_\chi=165$ MeV (see section \ref{sec4}).
This dependence on the chiral phase transition has to be examined more carefully in the future, which may have come from 
 the simplified sharp chiral phase transition we assumed.  Our objective
 in this work is to show that the CMW contribution to the slope $r$ can be order 1 comparable to the experiments.

\item For the CMW velocity $v_\chi$ inside the ``chiral boundary'', which depends on the inverse charge susceptibility, we use the lattice result for the susceptibility (at zero magnetic field) from Wuppertal Collaboration\cite{Borsanyi:2011sw}.
The $v_\chi$, as shown in (\ref{vchi}), 
is approximately linear in $eB$ for small $eB$, 
while for a sufficiently large $eB$ the susceptibility is modified such a way to saturate the causality bound $v_\chi\le1$.
This behavior has been verified previously in a holographic model \cite{Kharzeev:2010gd}.
We take this feature into account by replacing the naive $v_\chi$ obtained by using the lattice susceptibility (at zero magnetic field) with $v_\chi=1$ when the naive value exceeds the speed of light
\footnote{
In numerical calculations,
we set the largest value of $v_\chi$ to be $0.99$. Therefore all streamlines determined by \eq\eqref{CMWtraj} is time-like.
}.
The resulting shape of $v_\chi$ should be qualitatively same with the true behavior. 
Figure \ref{fig:vBvsT} shows the shape of $v_\chi$ as a function of temperature (the sudden drop is due to chiral transition at $T_\chi=150$ MeV).
This detail should not change our conclusion by order one factor. Future lattice study on the susceptibility in the presence of the magnetic field would be desirable to improve on this.
\begin{figure}[t]
	\centering
	\includegraphics[width=7cm]{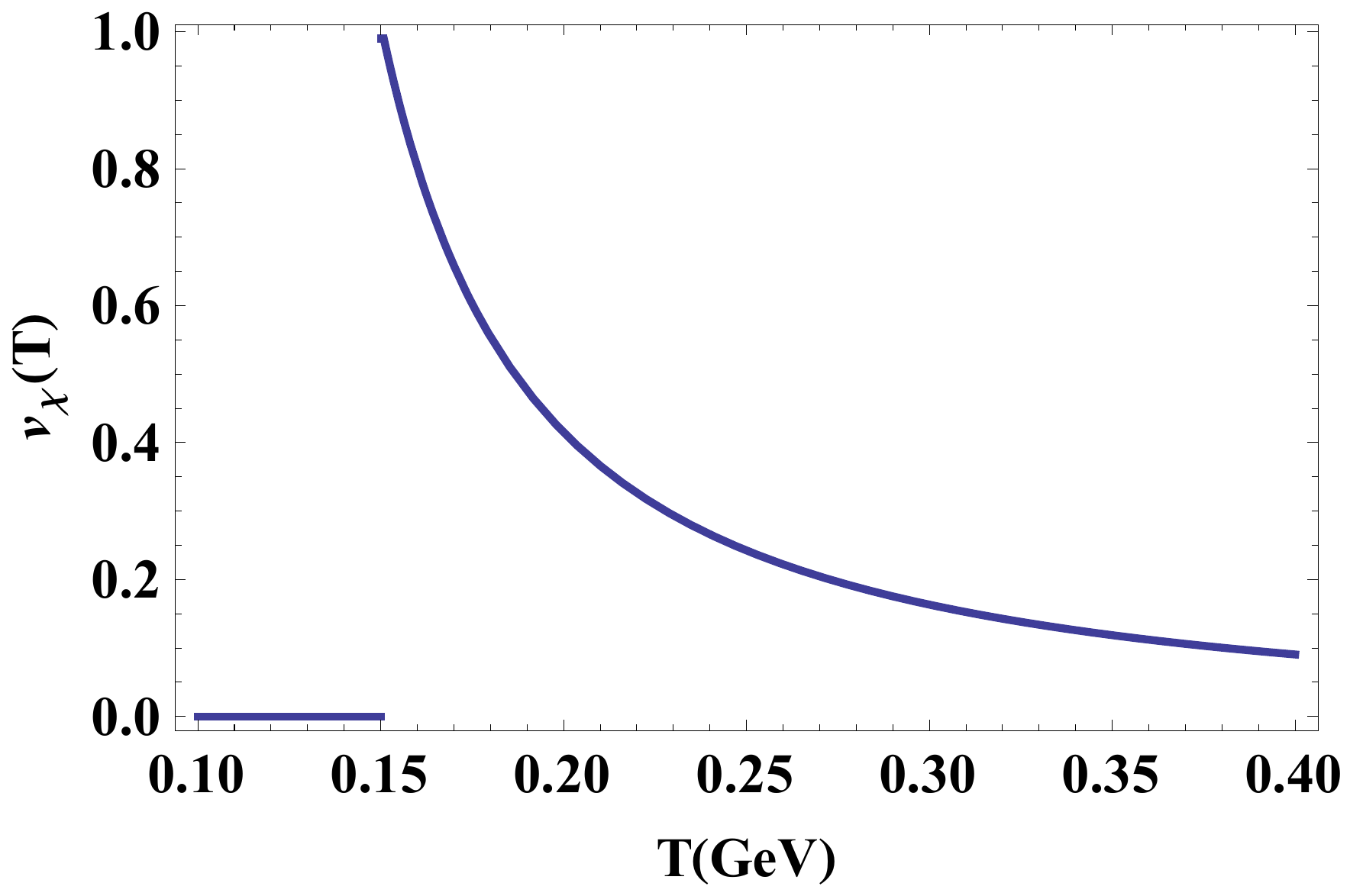}
		\caption{ (Color online) The CMW velocity $v_\chi$ with $eB/m^{2}_{\pi}=6$ as a function of temperature (in GeV) using the lattice result on susceptibility (with zero magnetic field) for a charge carrier with charge $q=(|q_u|+|q_d|)/2=1/2$. 
		The sudden drop at $T=150$ MeV is due to the chiral phase transition.	\label{fig:vBvsT}}
\end{figure}

\end{enumerate}

\section{
\label{sec3}
``Freezeout Hole Effect'' and chiral phase transition
}

\begin{figure}[t]
	\centering
	\includegraphics[width=7cm]{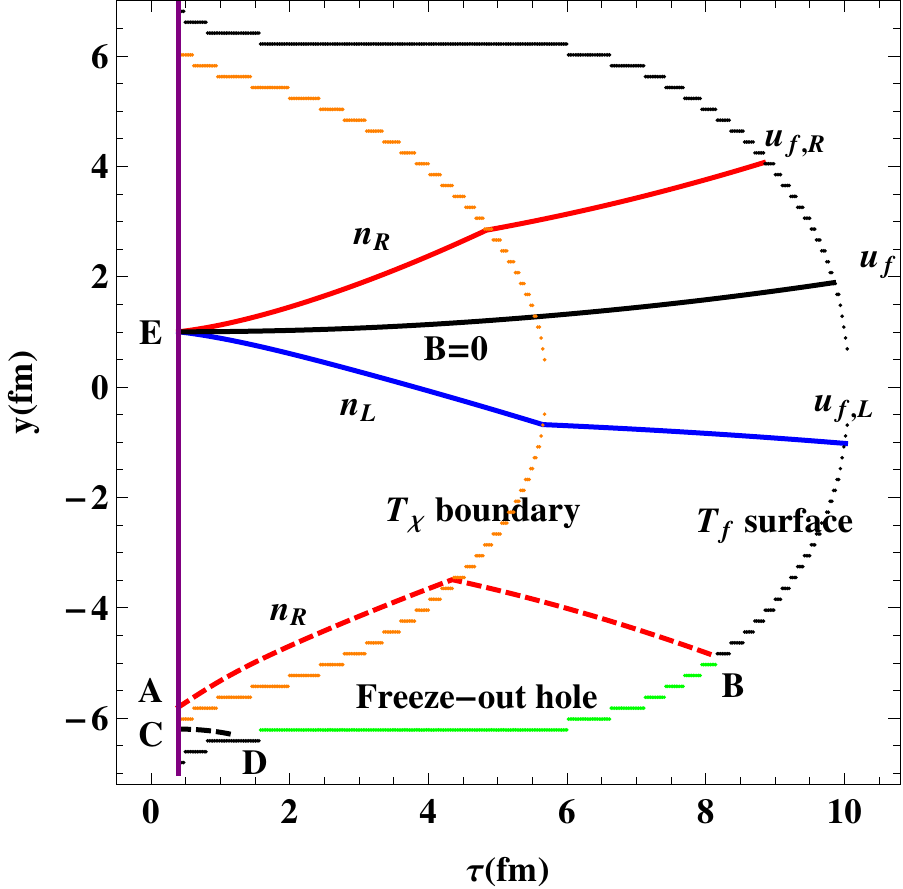}
		\caption{(Color online) Streamlines, freezeout surface and freeze-out hole(see text) projected on the plane $x=0.098$ fm for $b=7$ fm at RHIC energy. 
		The purple vertical line denotes the location of initial transverse plane $\tau_i =0.4$ fm.   	
		\label{fig:streamline}}
\end{figure}
As emphasized in the introduction, we identify 
a few important phenomena happening in the realistic simulation of the CMW 
that are responsible for a sizable contribution from the CMW to the slope parameter $r$: especially
the ``Freezeout Hole Effect''.
To highlight this effect, we show in Figure.~\ref{fig:streamline} some prototypical trajectories of the (right-handed chiral) charge cells in the $(\tau,y)$ plane, where $\tau$ is the Bjorken time and $y$ is the transverse coordinate along the magnetic field direction perpendicular to the reaction plane, starting from the initial surface at $\tau=\tau_i$ and ending on the freeze out surface (the curve on the far right). In the middle, we also point out the location of the chiral phase transition surface (the dotted orange curve) where the CMW velocity drops to zero. 

As seen in Figure.~\ref{fig:streamline},
the initial right-handed charge starting at the lowest position,
the point A,
in $y$ inside the chiral transition boundary
{(see orange curve)
 hits the freeze out surface at the point B after being affected by the CMW velocity.
The initial point slightly outside the chiral boundary (the point C) follows the background flow without CMW velocity, 
hitting the freeze out surface at the point D.
Since the trajectories can not intersect with each other as they are governed by the first order equation of motion, 
we conclude that the freeze out surface between the points B and D
does not encounter any right-handed chiral charges at all (see green curve connecting the points B and D)
: there is a ``Freezeout Hole'' with no right-handed chiral charge density upon which. 
For the left-handed charges, the freezeout hole exists in the mirror reflected place by $y\to-y$. 
This effect gives a sizable contribution to the momentum asymmetry with Cooper-Frye formula, 
leading to a dominant contribution from the CMW to the slope $r$.
We therefore emphasize that the proper treatment of the freeze out surface is crucially important in a trustable computation of the CMW contribution to the charge dependent elliptic flows of pions.

As a comparison, let us discuss what would be the case without the CMW.
Again, let us first consider the contribution to particle distribution in momentum space from one particular fluid cell which is located 
at initial transverse plane initially, say point E in Figure.~\ref{fig:streamline}.
In the absence of magnetic field,
the streamline of left-handed charge of that fluid cell coincides with that of right-handed charge ( see the black curve in Figure.~\ref{fig:streamline}).
When such streamline hits the freeze-out surface,
we denote the corresponding fluid velocity by $u^{\mu}_{f}$.
Similar to our previous formula \eq\eqref{eq:Nchcell},
we have
\begin{equation}
\label{eq:NchcellB0}
p^0{d^3 N^{\rm ch}_{\cell}\over d^3 p}\Bigg |_{B=0}=
{1\over (2\pi)^3}{A^i_{\rm cell}\tau_i \over T_f\chi_f}\, n_i\, \int d\xi \, 
 p_\mu \, u^\mu_{f} e^{-{p_{\mu} u^{\mu}_{f}\over T_f}}\, ,
\end{equation}
Moreover, 
due to the ideal hydrodynamic equation $\pd_{\mu}(s u^{\mu})=0$,
we also notice that along the streamline with respect to $u^{\mu}$,
total entropy inside the fluid cell is conserved.
Therefore we have $V^{i}_{\cell} s_{i} = V^{f}_{\cell} s_{f}$ and 
\begin{equation}
\label{eq:NbarcellB0}
p^0{d^3 \bar{N}_{\cell}\over d^3 p}\Bigg |_{B=0}=
{1\over (2\pi)^3}{A^i_{\rm cell}\tau_i\over T_f s_f} \, s_i \,\int d\xi \, 
 p_\mu \, u^\mu_{f} e^{-{p_{\mu} u^{\mu}_{f}\over T_f}}\, .
\end{equation}
Total $N^{\ch}$ and $\bar{N}$ can be determined by summing over all fluid cells as before.
Indeed, this is how we computed $\bar{v}_2$.
Comparing \eq\eqref{eq:NchcellB0} and \eq\eqref{eq:NbarcellB0},
we immediately see that if the charge distribution has a homogeneous profile of $n/s$
in the initial transverse space and there is no CMW, 
then $\bar v_2=v_2^{\rm ch}$ which implies that the slope parameter $r=2(\bar v_2-v_2^{\rm ch})$ vanishes. 
Therefore, we conclude that our reasonable initial condition and the freeze out condition would give zero (or negligibly small if the freeze out condition is not precisely the fixed temperature) contribution to the slope $r$,
if we did not include the CMW.  
We think that a large value of $r$ without the CMW found in Ref.~\cite{Hongo:2013cqa}
comes from their crude freeze-out surface at the constant ``time''.

\begin{figure}[t]
	\centering
	\includegraphics[width=7cm]{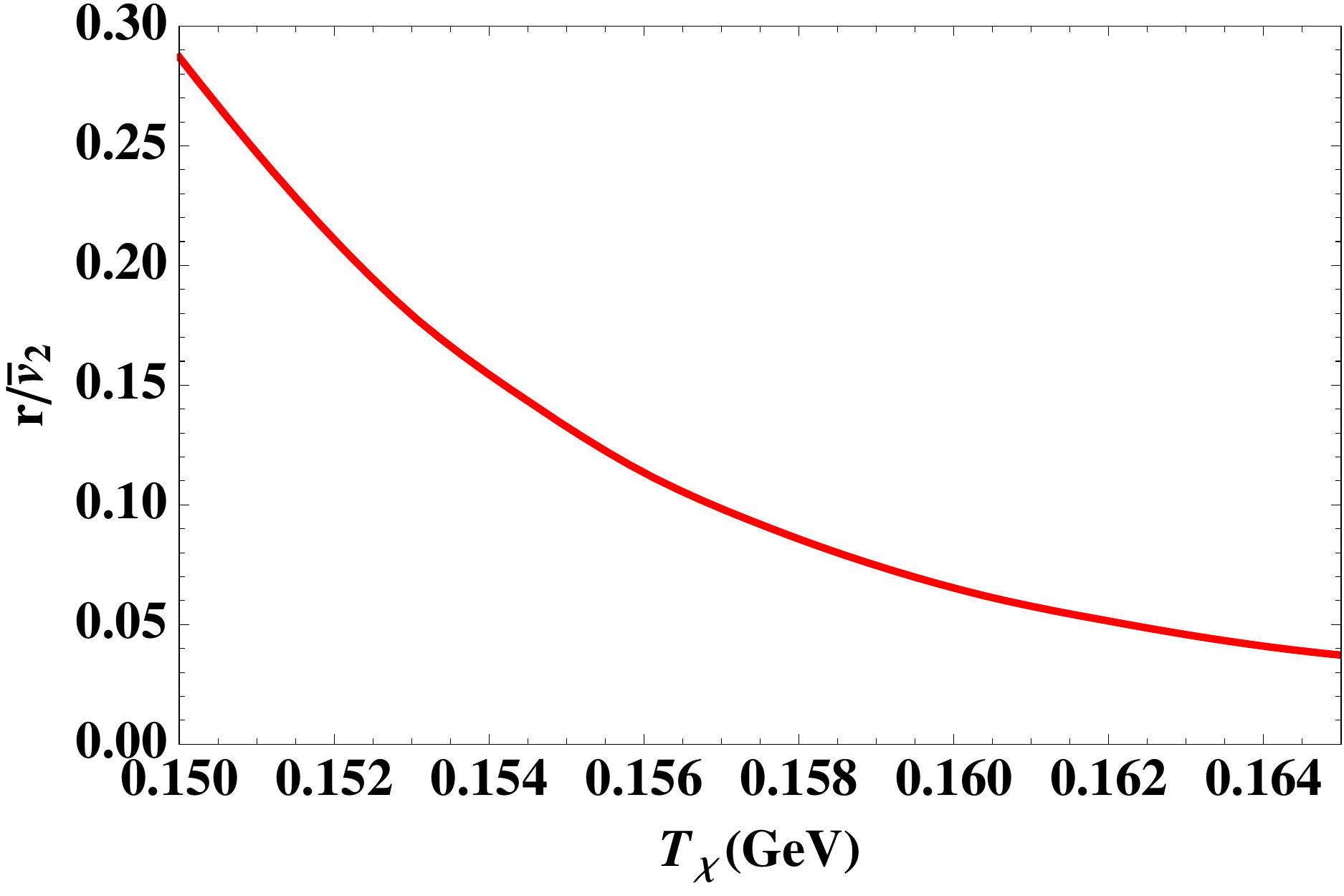}
		\caption{(Color online) The numerical results of $r/\bar v_2$ as a function of chiral phase transition temperature $T_\chi$ (in GeV) at RHIC energy with $b=7$~fm ({\it left}). 
The magnetic field is chosen to be $eB/m_\pi^2=6$ and $\tau_B=6$~fm.	\label{fig:Tchiral_dependence}}
\end{figure}

Another important ingredient of our simulation is the implementation of chiral phase transition (cross-over).
According to the recent lattice simulations \cite{Borsanyi:2011sw,Borsanyi:2010cj,Aoki:2009sc,*Aoki:2006br,*Aoki:2006we},
chiral phase cross-over happens in the region $T=0.15\sim 0.165$ GeV.
We present in Figure.~\ref{fig:Tchiral_dependence} the dependence of our numerical results on the chiral phase transition temperature at RHIC energy with $b=7$~fm, 
$eB/m_\pi^2=6$, and $\tau_B=6$~fm. The results of
$\kappa = r/\bar{v}_2$ is at $p_{\perp} = 0.8$~GeV. 
It is interesting to find a somewhat sensitive dependency on the chiral phase transition, 
dropping the results by a factor $5$ when we increase $T_\chi$ from $150$ MeV to $T_\chi=165$ MeV.
This can be attributed to the rapid decrease of $\chi$ which would in turn drastically increases the speed of CMW as shown in Figure.~\ref{fig:vBvsT}.
That effect suggests that CMW could be a useful probe of QCD chiral phase transition.

\section{Numerical Results and Comparison to Experiments \label{sec4} }

We now present the results of our numerical analysis. 
We have taken the charge of the charge carriers to be $q=(|q_{u}|+|q_{d}|)/2=1/2$.
For our optimistic scenario, we will use the value $T_\chi=150$ MeV in the following.
Before presenting our results as a function of impact parameter and the collision energy, we mention that our results also depend on a few more input parameters, such as
the strength ($eB/m_\pi^2$) and the lifetime ($\tau_B$) of the magnetic field.
The elliptic flow is also a function of the transverse momentum $p_\perp$ too.
Let us first discuss how our results are sensitive to these input parameters and the transverse momentum, and for this purpose let us look at RHIC energy $\sqrt{s_{NN}}=200$ GeV with the impact parameter $b=7$~fm.
Figure \ref{fig:tauBandB} shows $r/\bar v_2$ as a function of $p_\perp$ with $eB/m_\pi^2=6$
for different lifetimes $\tau_B=1,2,4,6$~fm. 
Perhaps, it is not surprising to find that the result is approximately linear in the lifetime $\tau_B$, 
but what is interesting here is that the ratio $r/\bar v_2$ is surprisingly constant over a wide range of transverse momentum $p_\perp$. 
That fact makes the ratio $\kappa=r/\bar{v}_2$ as a convenient parameter to characterize the contribution due to CMW to the slope $r$.
This is important since for low $p_\perp$, there are other sizable contributions such as resonance decays and the viscosity which we are neglecting. 
In the rest, we will present the results for a particular $p_\perp=0.8$ GeV bin where these additional effects can be neglected, 
but one can easily extrapolate our results for the CMW to a lower $p_\perp$ using the above constancy of $r/\bar v_2$.

\begin{figure}[t]
	\centering
	\includegraphics[width=0.45\textwidth]{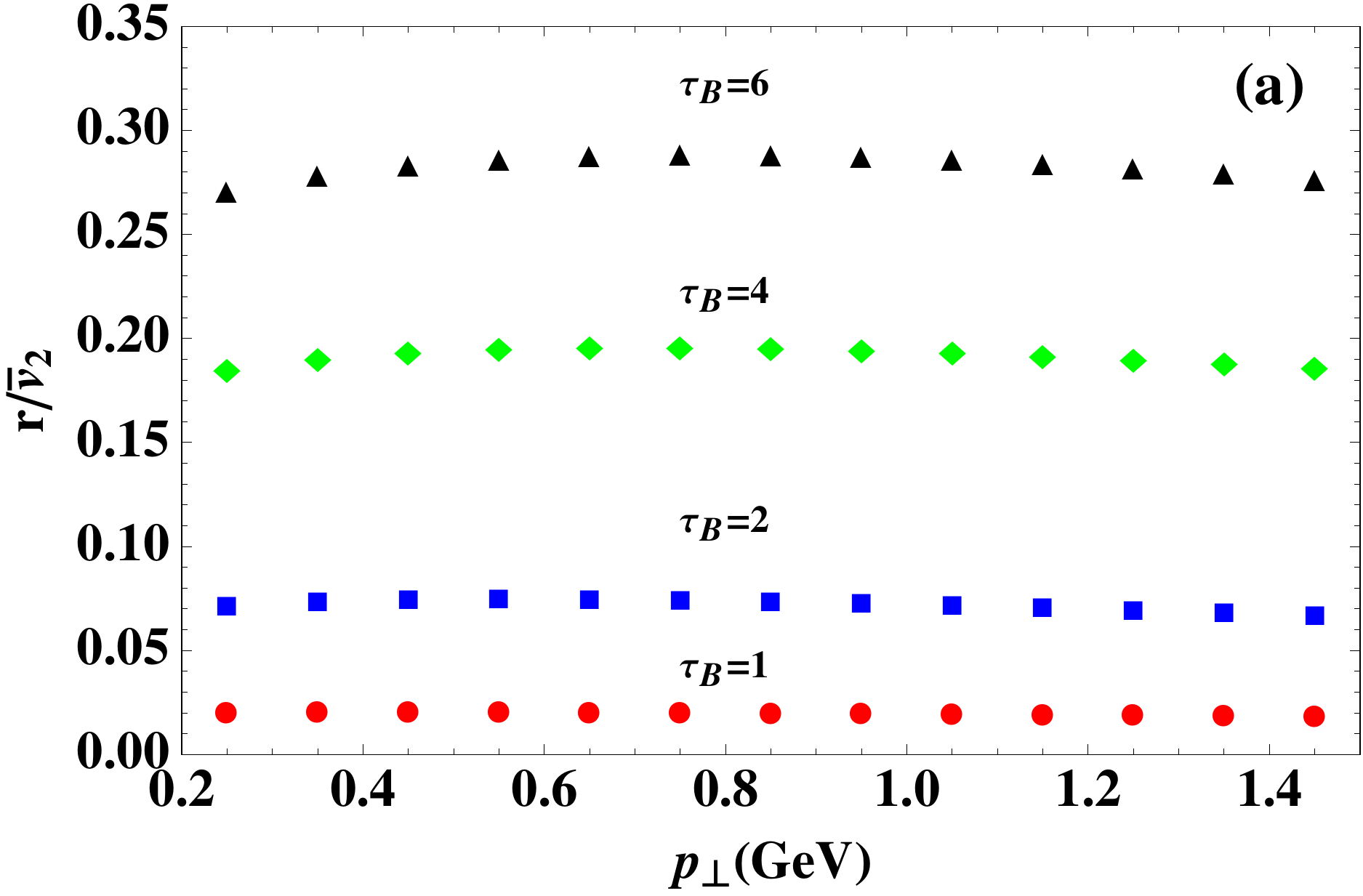}
	\includegraphics[width=0.45\textwidth]{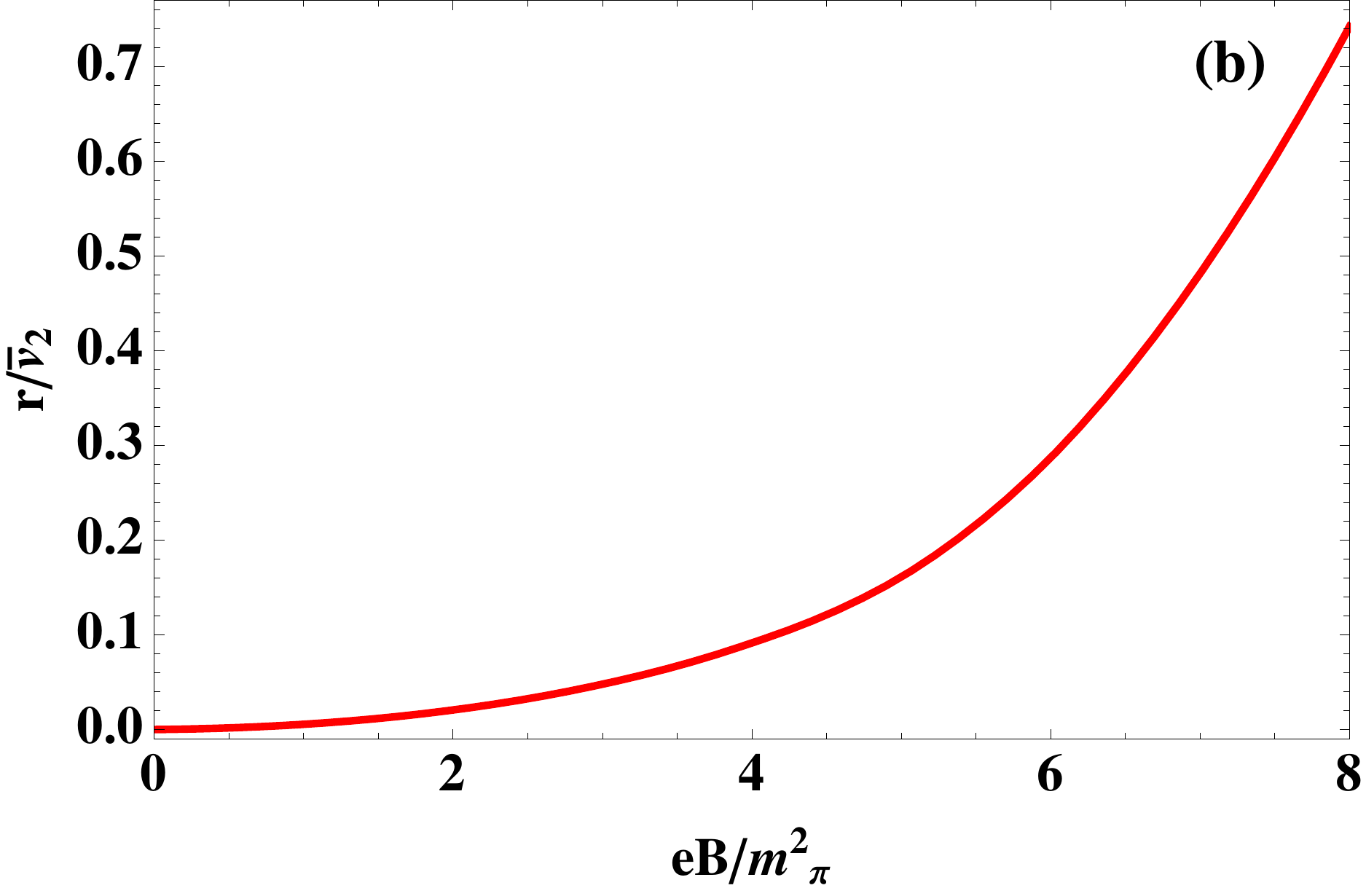}
		\caption{ (Color online)
		Left: The numerical results of $r/\bar v_2$ as a function of transverse momentum $p_{\perp}$(in GeV) for different lifetimes ($\tau_B=2,4,6$~fm) of the magnetic field at RHIC energy with $b=7$~fm. The magnetic field is chosen to be $eB/m_\pi^2=6$.
The right figure shows the results as a function of magnetic field with $\tau_B=6$~fm and $p_{\perp}=0.8$~GeV.	\label{fig:tauBandB}}
\end{figure}

The values of the magnetic field we will use for different impact parameters at the RHIC energy is in the table below 
based on the result in Ref.\cite{Bzdak:2011yy}.
\begin{table}[h]
\centering
\begin{tabular}{|l|l|l|l|}
\hline
b & $T_{i}$ at RHIC  & $T_{i}$ at LHC & $eB$ at RHIC \\
\hline
4 fm & 370 (MeV) & 470 (MeV)& \, $4 m^2_\pi$ \\
\hline
7 fm & 345 (MeV) & 440 (MeV) & \, $6 m^2_\pi$\\
\hline
10 fm & 270 (MeV) & 330 (MeV) & \, $7 m^2_\pi$ \\
\hline
\end{tabular}
\caption{\label{hydro-parameters}
Summary of parameters used for our CMW simulations for  different impact parameters and collision energies.
The $eB_{\rm{max}}$ at LHC at a given impact parameter would be given by $eB_{\rm{max}}$ of RHIC times $13.8$.}
\end{table} 
For the LHC energy, we scale up the magnetic field by $\gamma_{\rm LHC}/\gamma_{\rm RHIC}\approx 13.8$ and scale down the lifetime $\tau_B$ by the same factor.

Figure \ref{fig:RHIC} ({\it left}) shows our results of $r/\bar v_2$ at RHIC energy for three different impact parameters $b=4,7,10$~fm. 
For comparison, 
we plot the same quantity measured at RHIC experiments as a function of centrality(\textit{right})
\footnote{
It should be pointed out that in Ref.~\cite{Wang:2012qs,Ke:2012qb},
only experiment results of 
the slope $r$ for pions from $0.15$ GeV to $0.5$ GeV are given.
We estimate $\bar{v}_2$ in that $p_{\perp}$ range by
 $\bar{v}_2 \approx (\bar{v}_2(p_{\perp}=0.225GeV)+ 
\bar{v}_2(p_{\perp}=0.375GeV) +\bar{v}_2(p_{\perp}=0.525GeV))/3$ 
with $\bar{v}_2(p_{\perp})$ taken from Ref.~\cite{Adams:2004bi} to produce the right figure of Fig.~\ref{fig:RHIC}.
}.
 
We see that the contribution from the CMW we compute is comparable to the experimental values, 
somewhat less by a factor 2-3 for $\tau_B=6$~fm. 
This result is qualitatively similar to that in Refs.~\cite{Burnier:2011bf,Burnier:2012ae}. We also see that the impact parameter dependence, which is a non-trivial feature, qualitatively agrees with the experiments, showing a downfall behavior for a large impact parameter (note that the magnetic field for larger impact parameter is always bigger in our simulation). 
We are not claiming that the CMW contribution takes all account of the experimental value, as the slope $r$ is a P-even quantity subject to other possible effects, but we believe that our results point to that the CMW contribution is relevant and has to be taken into account in comparing to the experimental results.
\begin{figure}[t]
	\centering
	\includegraphics[width=7cm]{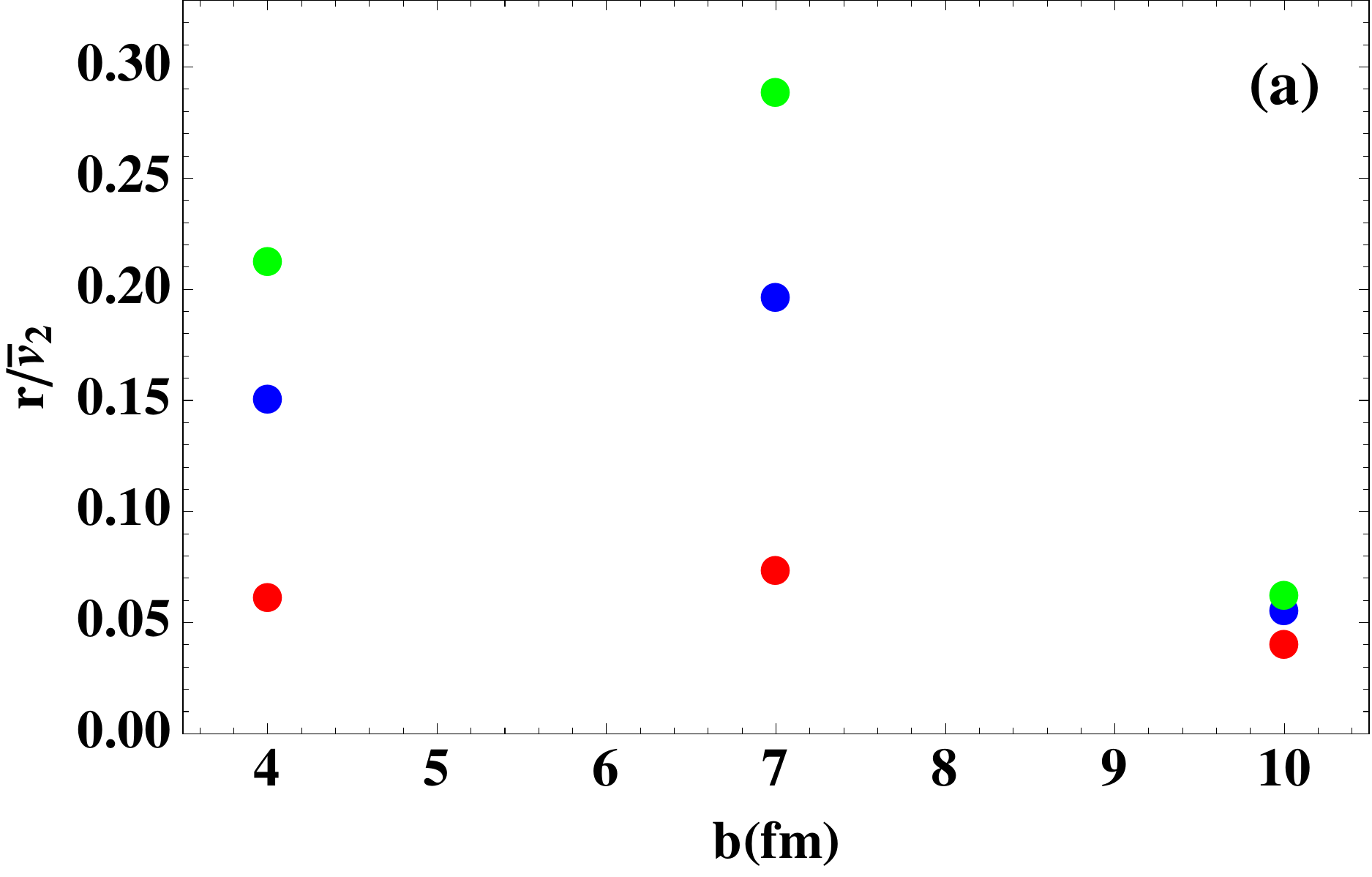}\includegraphics[width=7cm]{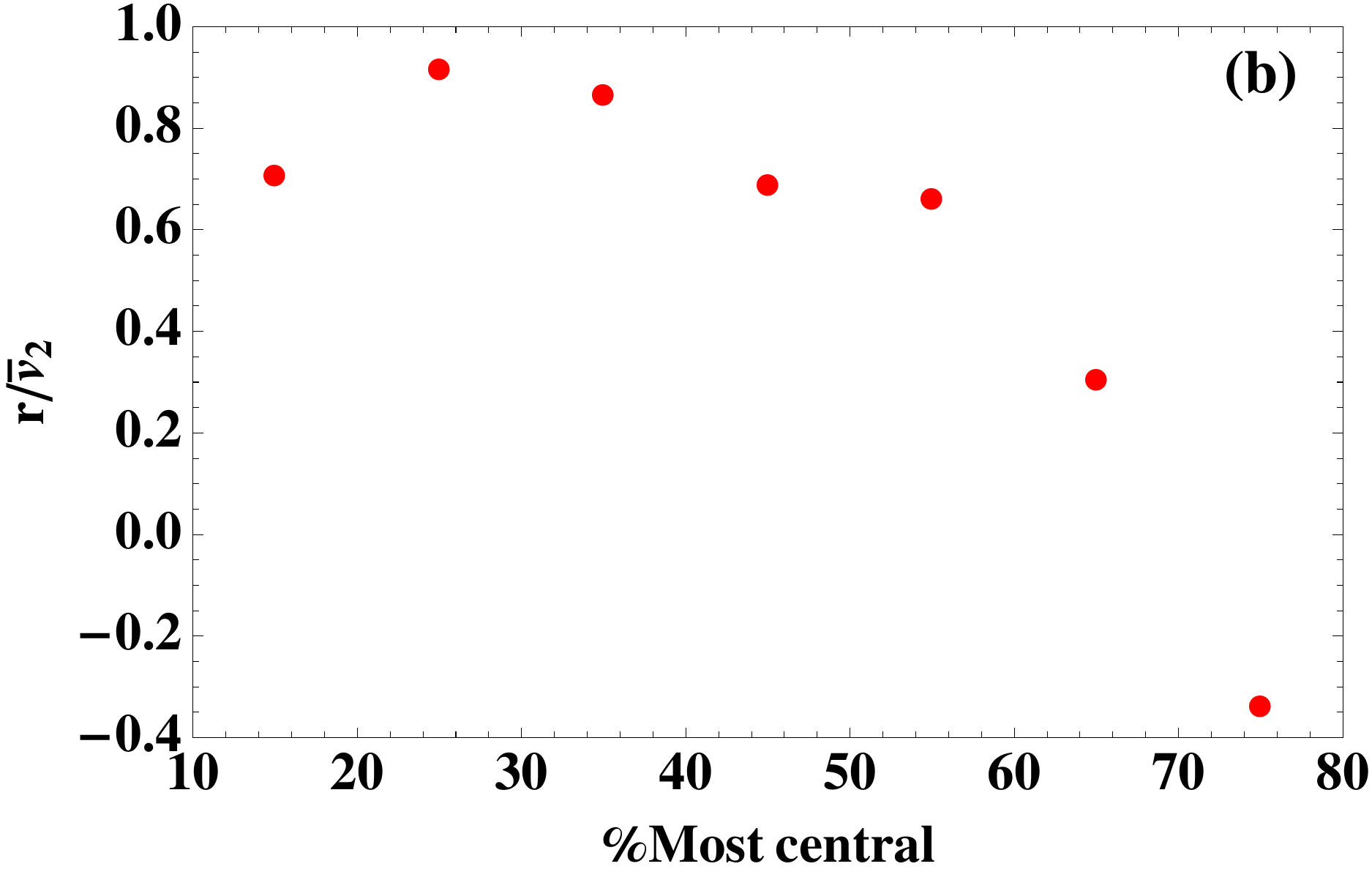}
		\caption{ (Color online) The numerical results of $r/\bar v_2$ as a function of the impact parameter at RHIC energy for three different lifetimes $\tau_B=2$ fm(red),$4$ fm(blue) $6$ fm(green) from bottom to top ({\it left}). The right figure shows the RHIC experiment results as a function of centrality.	\label{fig:RHIC}}
\end{figure}

In Figure.~\ref{fig:LHC} we show our prediction on $r/\bar v_2$ at the LHC energy 
for three different impact parameters, $b=4,7,10$~fm.
Remarkably,
the impact parameter dependence is different from that at the RHIC energy: there is no downfall behavior for a large impact parameter at the LHC energy.
Such difference between RHIC and LHC can be understood as follows.
It is seen from Figure.~\ref{fig:vBvsT} that at RHIC energy, 
the speed of CMW is small at high temperature or at early time.
However, at LHC where the magnetic field is much larger than that at RHIC,
the speed of CMW is already saturated at $v_\chi = 1$ at very early time. 
Consequently,
for sufficiently large magnetic field,
a large amount of charges would escape from the plasma due to CMW and freeze out at very early stage of the fireball expansion.
As radial flow has not yet developed at such early time,
the charge-dependent elliptic flow $v^{\ch}_{2}$ becomes very small.
As a result,
the ratio $\kappa = r/\bar v_2=2(\bar{v}_2 -v^{\ch}_2)/\bar{v}_2$ turns out to be larger in such situations. 
We emphasis here that the difference between RHIC and LHC that we discuss above would have been missed if the CMW equations were not solved in a realistic hydrodynamic background.

In the light of above discussion,
one may consider an extreme situation that all charges move to the freeze-out surface due to CMW before radial flow has been built up.
In such limit, $v^{\ch}_{2}\to 0$ and we obtain a theoretical upper bound to the ratio  
$\kappa = 2(\bar{v}_2 -v^{\ch}_2)/\bar{v}_2 \leq 2 $.
It would be interesting to see the results from LHC on $\kappa$ and compare it with the bound value $2$.
This comparison would provide an useful indicator of the CMW.

\begin{figure}[t]
	\centering
	\includegraphics[width=7cm]{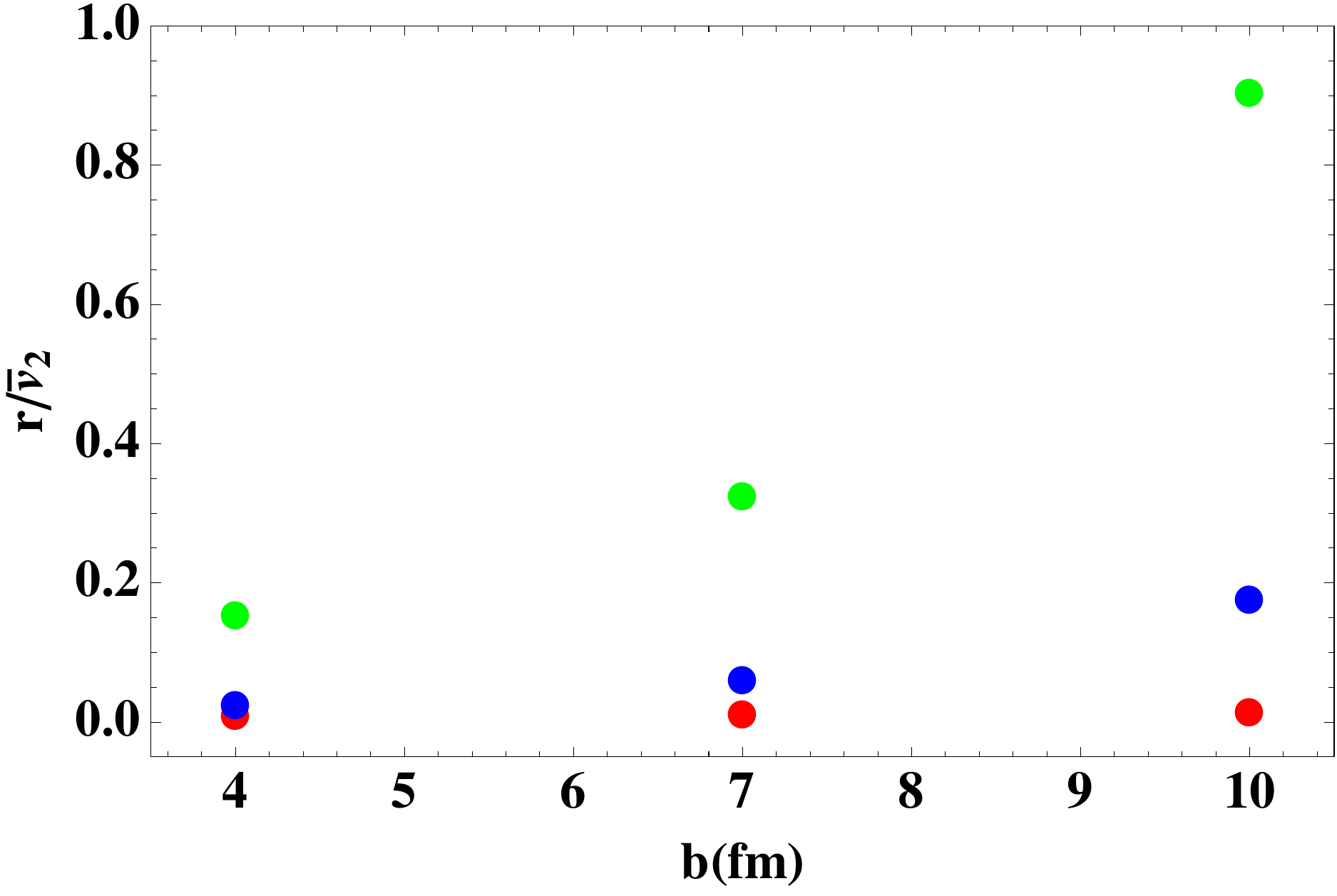}
		\caption{(Color online) The numerical results of $r/\bar v_2$ as a function of the impact parameter at LHC energy for three different lifetimes $\tau_B=2/13.8$ fm(red) $4/13.8$ fm (blue) $6/13.8$~fm(green) from bottom to top. 
We observe a very different impact parameter dependence than that at RHIC.\label{fig:LHC}}
\end{figure}

\section{Discussion}

In this work,
we study chiral magnetic wave (CMW) in realistic temperature and flow background. 
Our results are based on the realistic simulation of heavy-ion collisions 
and the proper initial/final states treatment points to that the CMW contribution to the charge dependent elliptic flow of pions is comparable to the experiments at RHIC.
We observe that the proper freeze out condition is crucial for a reliable computation of the CMW contributions to experimental observables. 
As the observable we are considering is subject to other background effects
(see for example Ref.~\cite{Bzdak:2013yla}),
we don't claim that the CMW contribution explains all portion of the data,. 
Our conclusion is that the CMW contribution is sizable and has to be taken into account.

Our result on the impact parameter dependence qualitatively agrees well with the experiments at RHIC. 
Our prediction for the LHC energy is quite different from that for RHIC, suggesting that we may be able to test the CMW contribution by looking at the impact parameter dependence at LHC energy.

We find somewhat large sensitivity of our results on the chiral phase transition temperature.
The lifetime of the magnetic field \cite{McLerran:2013hla,Tuchin:2013apa} is also a major source of uncertainty involved in our analysis. We haven't included fluctuations of the electromagnetic fields, and one should include viscous effects for a more precision computation of the observables. 
We leave all these important possible improvements to our future work.

As a final comment, it is known that the axial charge in the plasma is subject to decay via anomaly by the dynamical gauge fields.
One aspect of this is seen in the instability of the dynamical gauge fields in the presence of axial charge\cite{Joyce:1997uy,Akamatsu:2013pjd}.
In principle, this should be an important effect to be taken into account in realistic simulation of axial charge evolution.
We have neglected this since we are considering very small charge fluctuations (recall the charge asymmetry in FIG. \ref{fig1} is $A\sim 0.04$ maximum) and the resulting
instability time scale which is inversely proportional to the axial charge and $\alpha_{EM}$\cite{Akamatsu:2013pjd} is much larger than 10 fm. The situation for gluons should be addressed more carefully, which should be taken into account in the future analysis.

\acknowledgments

We thank Olga Evdokimov, Y. Hirono, M. Hongo, H. Ke, Dima Kharzeev, Jinfeng Liao, Misha Stephanov, Aihong Tang, and Gang Wang for helpful discussions and comments.
Y.Y. would like to thank Chun Shen for an e-mail correspondence on the initial parameters of the hydrodynamic simulation, Derek Teaney on discussion of freeze-out prescription and Todd Springer for suggestions on parametrizing lattice data. Y.Y. would extend his gratitude to the nuclear theory group of Stony Brook university for hospitality where part of this work has been done.
The work of Y.Y. is supported by the DOE grant No. DE-FG0201ER41195.

\bibliography{CMW}

\begin{thebibliography}{39}%
\makeatletter
\providecommand \@ifxundefined [1]{%
 \@ifx{#1\undefined}
}%
\providecommand \@ifnum [1]{%
 \ifnum #1\expandafter \@firstoftwo
 \else \expandafter \@secondoftwo
 \fi
}%
\providecommand \@ifx [1]{%
 \ifx #1\expandafter \@firstoftwo
 \else \expandafter \@secondoftwo
 \fi
}%
\providecommand \natexlab [1]{#1}%
\providecommand \enquote  [1]{``#1''}%
\providecommand \bibnamefont  [1]{#1}%
\providecommand \bibfnamefont [1]{#1}%
\providecommand \citenamefont [1]{#1}%
\providecommand \href@noop [0]{\@secondoftwo}%
\providecommand \href [0]{\begingroup \@sanitize@url \@href}%
\providecommand \@href[1]{\@@startlink{#1}\@@href}%
\providecommand \@@href[1]{\endgroup#1\@@endlink}%
\providecommand \@sanitize@url [0]{\catcode `\\12\catcode `\$12\catcode
  `\&12\catcode `\#12\catcode `\^12\catcode `\_12\catcode `\%12\relax}%
\providecommand \@@startlink[1]{}%
\providecommand \@@endlink[0]{}%
\providecommand \url  [0]{\begingroup\@sanitize@url \@url }%
\providecommand \@url [1]{\endgroup\@href {#1}{\urlprefix }}%
\providecommand \urlprefix  [0]{URL }%
\providecommand \Eprint [0]{\href }%
\providecommand \doibase [0]{http://dx.doi.org/}%
\providecommand \selectlanguage [0]{\@gobble}%
\providecommand \bibinfo  [0]{\@secondoftwo}%
\providecommand \bibfield  [0]{\@secondoftwo}%
\providecommand \translation [1]{[#1]}%
\providecommand \BibitemOpen [0]{}%
\providecommand \bibitemStop [0]{}%
\providecommand \bibitemNoStop [0]{.\EOS\space}%
\providecommand \EOS [0]{\spacefactor3000\relax}%
\providecommand \BibitemShut  [1]{\csname bibitem#1\endcsname}%
\let\auto@bib@innerbib\@empty
\bibitem [{\citenamefont {Kharzeev}\ \emph {et~al.}(2008)\citenamefont
  {Kharzeev}, \citenamefont {McLerran},\ and\ \citenamefont
  {Warringa}}]{Kharzeev:2007jp}%
  \BibitemOpen
  \bibfield  {author} {\bibinfo {author} {\bibfnamefont {D.~E.}\ \bibnamefont
  {Kharzeev}}, \bibinfo {author} {\bibfnamefont {L.~D.}\ \bibnamefont
  {McLerran}}, \ and\ \bibinfo {author} {\bibfnamefont {H.~J.}\ \bibnamefont
  {Warringa}},\ }\href {\doibase 10.1016/j.nuclphysa.2008.02.298} {\bibfield
  {journal} {\bibinfo  {journal} {Nucl.Phys.}\ }\textbf {\bibinfo {volume}
  {A803}},\ \bibinfo {pages} {227} (\bibinfo {year} {2008})},\ \Eprint
  {http://arxiv.org/abs/0711.0950} {arXiv:0711.0950 [hep-ph]} \BibitemShut
  {NoStop}%
\bibitem [{\citenamefont {Fukushima}\ \emph {et~al.}(2008)\citenamefont
  {Fukushima}, \citenamefont {Kharzeev},\ and\ \citenamefont
  {Warringa}}]{Fukushima:2008xe}%
  \BibitemOpen
  \bibfield  {author} {\bibinfo {author} {\bibfnamefont {K.}~\bibnamefont
  {Fukushima}}, \bibinfo {author} {\bibfnamefont {D.~E.}\ \bibnamefont
  {Kharzeev}}, \ and\ \bibinfo {author} {\bibfnamefont {H.~J.}\ \bibnamefont
  {Warringa}},\ }\href {\doibase 10.1103/PhysRevD.78.074033} {\bibfield
  {journal} {\bibinfo  {journal} {Phys. Rev. D}\ }\textbf {\bibinfo {volume}
  {78}},\ \bibinfo {pages} {074033} (\bibinfo {year} {2008})},\ \Eprint
  {http://arxiv.org/abs/0808.3382} {arXiv:0808.3382 [hep-ph]} \BibitemShut
  {NoStop}%
\bibitem [{\citenamefont {Son}\ and\ \citenamefont
  {Zhitnitsky}(2004)}]{Son:2004tq}%
  \BibitemOpen
  \bibfield  {author} {\bibinfo {author} {\bibfnamefont {D.}~\bibnamefont
  {Son}}\ and\ \bibinfo {author} {\bibfnamefont {A.~R.}\ \bibnamefont
  {Zhitnitsky}},\ }\href {\doibase 10.1103/PhysRevD.70.074018} {\bibfield
  {journal} {\bibinfo  {journal} {Phys.Rev.}\ }\textbf {\bibinfo {volume}
  {D70}},\ \bibinfo {pages} {074018} (\bibinfo {year} {2004})},\ \Eprint
  {http://arxiv.org/abs/hep-ph/0405216} {arXiv:hep-ph/0405216 [hep-ph]}
  \BibitemShut {NoStop}%
\bibitem [{\citenamefont {Son}\ and\ \citenamefont
  {Surowka}(2009)}]{Son:2009tf}%
  \BibitemOpen
  \bibfield  {author} {\bibinfo {author} {\bibfnamefont {D.~T.}\ \bibnamefont
  {Son}}\ and\ \bibinfo {author} {\bibfnamefont {P.}~\bibnamefont {Surowka}},\
  }\href {\doibase 10.1103/PhysRevLett.103.191601} {\bibfield  {journal}
  {\bibinfo  {journal} {Phys. Rev. Lett.}\ }\textbf {\bibinfo {volume} {103}},\
  \bibinfo {pages} {191601} (\bibinfo {year} {2009})},\ \Eprint
  {http://arxiv.org/abs/0906.5044} {arXiv:0906.5044 [hep-th]} \BibitemShut
  {NoStop}%
\bibitem [{\citenamefont {Voloshin}(2004)}]{Voloshin:2004vk}%
  \BibitemOpen
  \bibfield  {author} {\bibinfo {author} {\bibfnamefont {S.~A.}\ \bibnamefont
  {Voloshin}},\ }\href {\doibase 10.1103/PhysRevC.70.057901} {\bibfield
  {journal} {\bibinfo  {journal} {Phys.Rev.}\ }\textbf {\bibinfo {volume}
  {C70}},\ \bibinfo {pages} {057901} (\bibinfo {year} {2004})},\ \Eprint
  {http://arxiv.org/abs/hep-ph/0406311} {arXiv:hep-ph/0406311 [hep-ph]}
  \BibitemShut {NoStop}%
\bibitem [{\citenamefont {Abelev}\ \emph {et~al.}(2009)\citenamefont {Abelev}
  \emph {et~al.}}]{Abelev:2009ac}%
  \BibitemOpen
  \bibfield  {author} {\bibinfo {author} {\bibfnamefont {B.}~\bibnamefont
  {Abelev}} \emph {et~al.} (\bibinfo {collaboration} {STAR Collaboration}),\
  }\href {\doibase 10.1103/PhysRevLett.103.251601} {\bibfield  {journal}
  {\bibinfo  {journal} {Phys.Rev.Lett.}\ }\textbf {\bibinfo {volume} {103}},\
  \bibinfo {pages} {251601} (\bibinfo {year} {2009})},\ \Eprint
  {http://arxiv.org/abs/0909.1739} {arXiv:0909.1739 [nucl-ex]} \BibitemShut
  {NoStop}%
\bibitem [{\citenamefont {Selyuzhenkov}(2012)}]{Selyuzhenkov:2011xq}%
  \BibitemOpen
  \bibfield  {author} {\bibinfo {author} {\bibfnamefont {I.}~\bibnamefont
  {Selyuzhenkov}} (\bibinfo {collaboration} {ALICE Collaboration}),\ }\href
  {\doibase 10.1143/PTPS.193.153} {\bibfield  {journal} {\bibinfo  {journal}
  {Prog.Theor.Phys.Suppl.}\ }\textbf {\bibinfo {volume} {193}},\ \bibinfo
  {pages} {153} (\bibinfo {year} {2012})},\ \Eprint
  {http://arxiv.org/abs/1111.1875} {arXiv:1111.1875 [nucl-ex]} \BibitemShut
  {NoStop}%
\bibitem [{\citenamefont {Bzdak}\ \emph {et~al.}(2010)\citenamefont {Bzdak},
  \citenamefont {Koch},\ and\ \citenamefont {Liao}}]{Bzdak:2009fc}%
  \BibitemOpen
  \bibfield  {author} {\bibinfo {author} {\bibfnamefont {A.}~\bibnamefont
  {Bzdak}}, \bibinfo {author} {\bibfnamefont {V.}~\bibnamefont {Koch}}, \ and\
  \bibinfo {author} {\bibfnamefont {J.}~\bibnamefont {Liao}},\ }\href {\doibase
  10.1103/PhysRevC.81.031901} {\bibfield  {journal} {\bibinfo  {journal}
  {Phys.Rev.}\ }\textbf {\bibinfo {volume} {C81}},\ \bibinfo {pages} {031901}
  (\bibinfo {year} {2010})},\ \Eprint {http://arxiv.org/abs/0912.5050}
  {arXiv:0912.5050 [nucl-th]} \BibitemShut {NoStop}%
\bibitem [{\citenamefont {Wang}(2010)}]{Wang:2009kd}%
  \BibitemOpen
  \bibfield  {author} {\bibinfo {author} {\bibfnamefont {F.}~\bibnamefont
  {Wang}},\ }\href {\doibase 10.1103/PhysRevC.81.064902} {\bibfield  {journal}
  {\bibinfo  {journal} {Phys.Rev.}\ }\textbf {\bibinfo {volume} {C81}},\
  \bibinfo {pages} {064902} (\bibinfo {year} {2010})},\ \Eprint
  {http://arxiv.org/abs/0911.1482} {arXiv:0911.1482 [nucl-ex]} \BibitemShut
  {NoStop}%
\bibitem [{\citenamefont {Asakawa}\ \emph {et~al.}(2010)\citenamefont
  {Asakawa}, \citenamefont {Majumder},\ and\ \citenamefont
  {Muller}}]{Asakawa:2010bu}%
  \BibitemOpen
  \bibfield  {author} {\bibinfo {author} {\bibfnamefont {M.}~\bibnamefont
  {Asakawa}}, \bibinfo {author} {\bibfnamefont {A.}~\bibnamefont {Majumder}}, \
  and\ \bibinfo {author} {\bibfnamefont {B.}~\bibnamefont {Muller}},\ }\href
  {\doibase 10.1103/PhysRevC.81.064912} {\bibfield  {journal} {\bibinfo
  {journal} {Phys.Rev.}\ }\textbf {\bibinfo {volume} {C81}},\ \bibinfo {pages}
  {064912} (\bibinfo {year} {2010})},\ \Eprint {http://arxiv.org/abs/1003.2436}
  {arXiv:1003.2436 [hep-ph]} \BibitemShut {NoStop}%
\bibitem [{\citenamefont {Pratt}\ \emph {et~al.}(2011)\citenamefont {Pratt},
  \citenamefont {Schlichting},\ and\ \citenamefont {Gavin}}]{Pratt:2010zn}%
  \BibitemOpen
  \bibfield  {author} {\bibinfo {author} {\bibfnamefont {S.}~\bibnamefont
  {Pratt}}, \bibinfo {author} {\bibfnamefont {S.}~\bibnamefont {Schlichting}},
  \ and\ \bibinfo {author} {\bibfnamefont {S.}~\bibnamefont {Gavin}},\ }\href
  {\doibase 10.1103/PhysRevC.84.024909} {\bibfield  {journal} {\bibinfo
  {journal} {Phys.Rev.}\ }\textbf {\bibinfo {volume} {C84}},\ \bibinfo {pages}
  {024909} (\bibinfo {year} {2011})},\ \Eprint {http://arxiv.org/abs/1011.6053}
  {arXiv:1011.6053 [nucl-th]} \BibitemShut {NoStop}%
\bibitem [{\citenamefont {Bzdak}\ \emph {et~al.}(2013)\citenamefont {Bzdak},
  \citenamefont {Koch},\ and\ \citenamefont {Liao}}]{Bzdak:2012ia}%
  \BibitemOpen
  \bibfield  {author} {\bibinfo {author} {\bibfnamefont {A.}~\bibnamefont
  {Bzdak}}, \bibinfo {author} {\bibfnamefont {V.}~\bibnamefont {Koch}}, \ and\
  \bibinfo {author} {\bibfnamefont {J.}~\bibnamefont {Liao}},\ }\href {\doibase
  10.1007/978-3-642-37305-3_19} {\bibfield  {journal} {\bibinfo  {journal}
  {Lect.Notes Phys.}\ }\textbf {\bibinfo {volume} {871}},\ \bibinfo {pages}
  {503} (\bibinfo {year} {2013})},\ \Eprint {http://arxiv.org/abs/1207.7327}
  {arXiv:1207.7327} \BibitemShut {NoStop}%
\bibitem [{\citenamefont {Kharzeev}\ and\ \citenamefont
  {Yee}(2011)}]{Kharzeev:2010gd}%
  \BibitemOpen
  \bibfield  {author} {\bibinfo {author} {\bibfnamefont {D.~E.}\ \bibnamefont
  {Kharzeev}}\ and\ \bibinfo {author} {\bibfnamefont {H.-U.}\ \bibnamefont
  {Yee}},\ }\href {\doibase 10.1103/PhysRevD.83.085007} {\bibfield  {journal}
  {\bibinfo  {journal} {Phys.Rev.}\ }\textbf {\bibinfo {volume} {D83}},\
  \bibinfo {pages} {085007} (\bibinfo {year} {2011})},\ \Eprint
  {http://arxiv.org/abs/1012.6026} {arXiv:1012.6026 [hep-th]} \BibitemShut
  {NoStop}%
\bibitem [{\citenamefont {Newman}(2006)}]{Newman:2005hd}%
  \BibitemOpen
  \bibfield  {author} {\bibinfo {author} {\bibfnamefont {G.}~\bibnamefont
  {Newman}},\ }\href {\doibase 10.1088/1126-6708/2006/01/158} {\bibfield
  {journal} {\bibinfo  {journal} {JHEP}\ }\textbf {\bibinfo {volume} {0601}},\
  \bibinfo {pages} {158} (\bibinfo {year} {2006})},\ \Eprint
  {http://arxiv.org/abs/hep-ph/0511236} {arXiv:hep-ph/0511236 [hep-ph]}
  \BibitemShut {NoStop}%
\bibitem [{\citenamefont {Wang}(2013)}]{Wang:2012qs}%
  \BibitemOpen
  \bibfield  {author} {\bibinfo {author} {\bibfnamefont {G.}~\bibnamefont
  {Wang}} (\bibinfo {collaboration} {STAR Collaboration}),\ }\href {\doibase
  10.1016/j.nuclphysa.2013.01.069} {\bibfield  {journal} {\bibinfo  {journal}
  {Nucl.Phys.A904-905}\ }\textbf {\bibinfo {volume} {2013}},\ \bibinfo {pages}
  {248c} (\bibinfo {year} {2013})},\ \Eprint {http://arxiv.org/abs/1210.5498}
  {arXiv:1210.5498 [nucl-ex]} \BibitemShut {NoStop}%
\bibitem [{\citenamefont {Ke}(2012)}]{Ke:2012qb}%
  \BibitemOpen
  \bibfield  {author} {\bibinfo {author} {\bibfnamefont {H.}~\bibnamefont {Ke}}
  (\bibinfo {collaboration} {STAR Collaboration}),\ }\href {\doibase
  10.1088/1742-6596/389/1/012035} {\bibfield  {journal} {\bibinfo  {journal}
  {J.Phys.Conf.Ser.}\ }\textbf {\bibinfo {volume} {389}},\ \bibinfo {pages}
  {012035} (\bibinfo {year} {2012})},\ \Eprint {http://arxiv.org/abs/1211.3216}
  {arXiv:1211.3216 [nucl-ex]} \BibitemShut {NoStop}%
\bibitem [{\citenamefont {Stephanov}\ and\ \citenamefont
  {Yee}(2013)}]{Stephanov:2013tga}%
  \BibitemOpen
  \bibfield  {author} {\bibinfo {author} {\bibfnamefont {M.}~\bibnamefont
  {Stephanov}}\ and\ \bibinfo {author} {\bibfnamefont {H.-U.}\ \bibnamefont
  {Yee}},\ }\href@noop {} {\bibfield  {journal} {\bibinfo  {journal}
  {Phys.Rev.}\ }\textbf {\bibinfo {volume} {C88}},\ \bibinfo {pages} {014908}
  (\bibinfo {year} {2013})},\ \Eprint {http://arxiv.org/abs/1304.6410}
  {arXiv:1304.6410 [nucl-th]} \BibitemShut {NoStop}%
\bibitem [{\citenamefont {Burnier}\ \emph {et~al.}(2011)\citenamefont
  {Burnier}, \citenamefont {Kharzeev}, \citenamefont {Liao},\ and\
  \citenamefont {Yee}}]{Burnier:2011bf}%
  \BibitemOpen
  \bibfield  {author} {\bibinfo {author} {\bibfnamefont {Y.}~\bibnamefont
  {Burnier}}, \bibinfo {author} {\bibfnamefont {D.~E.}\ \bibnamefont
  {Kharzeev}}, \bibinfo {author} {\bibfnamefont {J.}~\bibnamefont {Liao}}, \
  and\ \bibinfo {author} {\bibfnamefont {H.-U.}\ \bibnamefont {Yee}},\ }\href
  {\doibase 10.1103/PhysRevLett.107.052303} {\bibfield  {journal} {\bibinfo
  {journal} {Phys.Rev.Lett.}\ }\textbf {\bibinfo {volume} {107}},\ \bibinfo
  {pages} {052303} (\bibinfo {year} {2011})},\ \Eprint
  {http://arxiv.org/abs/1103.1307} {arXiv:1103.1307 [hep-ph]} \BibitemShut
  {NoStop}%
\bibitem [{\citenamefont {Burnier}\ \emph {et~al.}(2012)\citenamefont
  {Burnier}, \citenamefont {Kharzeev}, \citenamefont {Liao},\ and\
  \citenamefont {Yee}}]{Burnier:2012ae}%
  \BibitemOpen
  \bibfield  {author} {\bibinfo {author} {\bibfnamefont {Y.}~\bibnamefont
  {Burnier}}, \bibinfo {author} {\bibfnamefont {D.}~\bibnamefont {Kharzeev}},
  \bibinfo {author} {\bibfnamefont {J.}~\bibnamefont {Liao}}, \ and\ \bibinfo
  {author} {\bibfnamefont {H.-U.}\ \bibnamefont {Yee}},\ }\href@noop {} {\
  (\bibinfo {year} {2012})},\ \Eprint {http://arxiv.org/abs/1208.2537}
  {arXiv:1208.2537 [hep-ph]} \BibitemShut {NoStop}%
\bibitem [{\citenamefont {Gorbar}\ \emph {et~al.}(2011)\citenamefont {Gorbar},
  \citenamefont {Miransky},\ and\ \citenamefont {Shovkovy}}]{Gorbar:2011ya}%
  \BibitemOpen
  \bibfield  {author} {\bibinfo {author} {\bibfnamefont {E.}~\bibnamefont
  {Gorbar}}, \bibinfo {author} {\bibfnamefont {V.}~\bibnamefont {Miransky}}, \
  and\ \bibinfo {author} {\bibfnamefont {I.}~\bibnamefont {Shovkovy}},\ }\href
  {\doibase 10.1103/PhysRevD.83.085003} {\bibfield  {journal} {\bibinfo
  {journal} {Phys.Rev.}\ }\textbf {\bibinfo {volume} {D83}},\ \bibinfo {pages}
  {085003} (\bibinfo {year} {2011})},\ \Eprint {http://arxiv.org/abs/1101.4954}
  {arXiv:1101.4954 [hep-ph]} \BibitemShut {NoStop}%
\bibitem [{\citenamefont {Hongo}\ \emph {et~al.}(2013)\citenamefont {Hongo},
  \citenamefont {Hirono},\ and\ \citenamefont {Hirano}}]{Hongo:2013cqa}%
  \BibitemOpen
  \bibfield  {author} {\bibinfo {author} {\bibfnamefont {M.}~\bibnamefont
  {Hongo}}, \bibinfo {author} {\bibfnamefont {Y.}~\bibnamefont {Hirono}}, \
  and\ \bibinfo {author} {\bibfnamefont {T.}~\bibnamefont {Hirano}},\
  }\href@noop {} {\  (\bibinfo {year} {2013})},\ \Eprint
  {http://arxiv.org/abs/1309.2823} {arXiv:1309.2823 [nucl-th]} \BibitemShut
  {NoStop}%
\bibitem [{\citenamefont {Taghavi}\ and\ \citenamefont
  {Wiedemann}(2013)}]{Taghavi:2013ena}%
  \BibitemOpen
  \bibfield  {author} {\bibinfo {author} {\bibfnamefont {S.~F.}\ \bibnamefont
  {Taghavi}}\ and\ \bibinfo {author} {\bibfnamefont {U.~A.}\ \bibnamefont
  {Wiedemann}},\ }\href@noop {} {\  (\bibinfo {year} {2013})},\ \Eprint
  {http://arxiv.org/abs/1310.0193} {arXiv:1310.0193 [hep-ph]} \BibitemShut
  {NoStop}%
\bibitem [{\citenamefont {Borsanyi}\ \emph {et~al.}(2010)\citenamefont
  {Borsanyi}, \citenamefont {Endrodi}, \citenamefont {Fodor}, \citenamefont
  {Jakovac}, \citenamefont {Katz} \emph {et~al.}}]{Borsanyi:2010cj}%
  \BibitemOpen
  \bibfield  {author} {\bibinfo {author} {\bibfnamefont {S.}~\bibnamefont
  {Borsanyi}}, \bibinfo {author} {\bibfnamefont {G.}~\bibnamefont {Endrodi}},
  \bibinfo {author} {\bibfnamefont {Z.}~\bibnamefont {Fodor}}, \bibinfo
  {author} {\bibfnamefont {A.}~\bibnamefont {Jakovac}}, \bibinfo {author}
  {\bibfnamefont {S.~D.}\ \bibnamefont {Katz}},  \emph {et~al.},\ }\href
  {\doibase 10.1007/JHEP11(2010)077} {\bibfield  {journal} {\bibinfo  {journal}
  {JHEP}\ }\textbf {\bibinfo {volume} {1011}},\ \bibinfo {pages} {077}
  (\bibinfo {year} {2010})},\ \Eprint {http://arxiv.org/abs/1007.2580}
  {arXiv:1007.2580 [hep-lat]} \BibitemShut {NoStop}%
\bibitem [{\citenamefont {Borsanyi}\ \emph {et~al.}(2012)\citenamefont
  {Borsanyi}, \citenamefont {Fodor}, \citenamefont {Katz}, \citenamefont
  {Krieg}, \citenamefont {Ratti} \emph {et~al.}}]{Borsanyi:2011sw}%
  \BibitemOpen
  \bibfield  {author} {\bibinfo {author} {\bibfnamefont {S.}~\bibnamefont
  {Borsanyi}}, \bibinfo {author} {\bibfnamefont {Z.}~\bibnamefont {Fodor}},
  \bibinfo {author} {\bibfnamefont {S.~D.}\ \bibnamefont {Katz}}, \bibinfo
  {author} {\bibfnamefont {S.}~\bibnamefont {Krieg}}, \bibinfo {author}
  {\bibfnamefont {C.}~\bibnamefont {Ratti}},  \emph {et~al.},\ }\href {\doibase
  10.1007/JHEP01(2012)138} {\bibfield  {journal} {\bibinfo  {journal} {JHEP}\
  }\textbf {\bibinfo {volume} {1201}},\ \bibinfo {pages} {138} (\bibinfo {year}
  {2012})},\ \Eprint {http://arxiv.org/abs/1112.4416} {arXiv:1112.4416
  [hep-lat]} \BibitemShut {NoStop}%
\bibitem [{\citenamefont {Romatschke}\ and\ \citenamefont
  {Romatschke}(2007)}]{Romatschke:2007mq}%
  \BibitemOpen
  \bibfield  {author} {\bibinfo {author} {\bibfnamefont {P.}~\bibnamefont
  {Romatschke}}\ and\ \bibinfo {author} {\bibfnamefont {U.}~\bibnamefont
  {Romatschke}},\ }\href {\doibase 10.1103/PhysRevLett.99.172301} {\bibfield
  {journal} {\bibinfo  {journal} {Phys.Rev.Lett.}\ }\textbf {\bibinfo {volume}
  {99}},\ \bibinfo {pages} {172301} (\bibinfo {year} {2007})},\ \Eprint
  {http://arxiv.org/abs/0706.1522} {arXiv:0706.1522 [nucl-th]} \BibitemShut
  {NoStop}%
\bibitem [{\citenamefont {Luzum}\ and\ \citenamefont
  {Romatschke}(2008)}]{Luzum:2008cw}%
  \BibitemOpen
  \bibfield  {author} {\bibinfo {author} {\bibfnamefont {M.}~\bibnamefont
  {Luzum}}\ and\ \bibinfo {author} {\bibfnamefont {P.}~\bibnamefont
  {Romatschke}},\ }\href {\doibase 10.1103/PhysRevC.78.034915,
  10.1103/PhysRevC.79.039903} {\bibfield  {journal} {\bibinfo  {journal}
  {Phys.Rev.}\ }\textbf {\bibinfo {volume} {C78}},\ \bibinfo {pages} {034915}
  (\bibinfo {year} {2008})},\ \Eprint {http://arxiv.org/abs/0804.4015}
  {arXiv:0804.4015 [nucl-th]} \BibitemShut {NoStop}%
\bibitem [{\citenamefont {Shen}\ \emph {et~al.}(2010)\citenamefont {Shen},
  \citenamefont {Heinz}, \citenamefont {Huovinen},\ and\ \citenamefont
  {Song}}]{Shen:2010uy}%
  \BibitemOpen
  \bibfield  {author} {\bibinfo {author} {\bibfnamefont {C.}~\bibnamefont
  {Shen}}, \bibinfo {author} {\bibfnamefont {U.}~\bibnamefont {Heinz}},
  \bibinfo {author} {\bibfnamefont {P.}~\bibnamefont {Huovinen}}, \ and\
  \bibinfo {author} {\bibfnamefont {H.}~\bibnamefont {Song}},\ }\href {\doibase
  10.1103/PhysRevC.82.054904} {\bibfield  {journal} {\bibinfo  {journal}
  {Phys.Rev.}\ }\textbf {\bibinfo {volume} {C82}},\ \bibinfo {pages} {054904}
  (\bibinfo {year} {2010})},\ \Eprint {http://arxiv.org/abs/1010.1856}
  {arXiv:1010.1856 [nucl-th]} \BibitemShut {NoStop}%
\bibitem [{\citenamefont {Basar}\ \emph {et~al.}(2012)\citenamefont {Basar},
  \citenamefont {Kharzeev}, \citenamefont {Kharzeev},\ and\ \citenamefont
  {Skokov}}]{Basar:2012bp}%
  \BibitemOpen
  \bibfield  {author} {\bibinfo {author} {\bibfnamefont {G.}~\bibnamefont
  {Basar}}, \bibinfo {author} {\bibfnamefont {D.}~\bibnamefont {Kharzeev}},
  \bibinfo {author} {\bibfnamefont {D.}~\bibnamefont {Kharzeev}}, \ and\
  \bibinfo {author} {\bibfnamefont {V.}~\bibnamefont {Skokov}},\ }\href
  {\doibase 10.1103/PhysRevLett.109.202303} {\bibfield  {journal} {\bibinfo
  {journal} {Phys.Rev.Lett.}\ }\textbf {\bibinfo {volume} {109}},\ \bibinfo
  {pages} {202303} (\bibinfo {year} {2012})},\ \Eprint
  {http://arxiv.org/abs/1206.1334} {arXiv:1206.1334 [hep-ph]} \BibitemShut
  {NoStop}%
\bibitem [{\citenamefont {Shen}\ \emph {et~al.}(2011)\citenamefont {Shen},
  \citenamefont {Heinz}, \citenamefont {Huovinen},\ and\ \citenamefont
  {Song}}]{Shen:2011eg}%
  \BibitemOpen
  \bibfield  {author} {\bibinfo {author} {\bibfnamefont {C.}~\bibnamefont
  {Shen}}, \bibinfo {author} {\bibfnamefont {U.}~\bibnamefont {Heinz}},
  \bibinfo {author} {\bibfnamefont {P.}~\bibnamefont {Huovinen}}, \ and\
  \bibinfo {author} {\bibfnamefont {H.}~\bibnamefont {Song}},\ }\href {\doibase
  10.1103/PhysRevC.84.044903} {\bibfield  {journal} {\bibinfo  {journal}
  {Phys.Rev.}\ }\textbf {\bibinfo {volume} {C84}},\ \bibinfo {pages} {044903}
  (\bibinfo {year} {2011})},\ \Eprint {http://arxiv.org/abs/1105.3226}
  {arXiv:1105.3226 [nucl-th]} \BibitemShut {NoStop}%
\bibitem [{\citenamefont {Adams}\ \emph {et~al.}(2005)\citenamefont {Adams}
  \emph {et~al.}}]{Adams:2004bi}%
  \BibitemOpen
  \bibfield  {author} {\bibinfo {author} {\bibfnamefont {J.}~\bibnamefont
  {Adams}} \emph {et~al.} (\bibinfo {collaboration} {STAR Collaboration}),\
  }\href {\doibase 10.1103/PhysRevC.72.014904} {\bibfield  {journal} {\bibinfo
  {journal} {Phys.Rev.}\ }\textbf {\bibinfo {volume} {C72}},\ \bibinfo {pages}
  {014904} (\bibinfo {year} {2005})},\ \Eprint
  {http://arxiv.org/abs/nucl-ex/0409033} {arXiv:nucl-ex/0409033 [nucl-ex]}
  \BibitemShut {NoStop}%
\bibitem [{\citenamefont {Aoki}\ \emph {et~al.}(2009)\citenamefont {Aoki},
  \citenamefont {Borsanyi}, \citenamefont {Durr}, \citenamefont {Fodor},
  \citenamefont {Katz} \emph {et~al.}}]{Aoki:2009sc}%
  \BibitemOpen
  \bibfield  {author} {\bibinfo {author} {\bibfnamefont {Y.}~\bibnamefont
  {Aoki}}, \bibinfo {author} {\bibfnamefont {S.}~\bibnamefont {Borsanyi}},
  \bibinfo {author} {\bibfnamefont {S.}~\bibnamefont {Durr}}, \bibinfo {author}
  {\bibfnamefont {Z.}~\bibnamefont {Fodor}}, \bibinfo {author} {\bibfnamefont
  {S.~D.}\ \bibnamefont {Katz}},  \emph {et~al.},\ }\href {\doibase
  10.1088/1126-6708/2009/06/088} {\bibfield  {journal} {\bibinfo  {journal}
  {JHEP}\ }\textbf {\bibinfo {volume} {0906}},\ \bibinfo {pages} {088}
  (\bibinfo {year} {2009})},\ \Eprint {http://arxiv.org/abs/0903.4155}
  {arXiv:0903.4155 [hep-lat]} \BibitemShut {NoStop}%
\bibitem [{\citenamefont {Aoki}\ \emph
  {et~al.}(2006{\natexlab{a}})\citenamefont {Aoki}, \citenamefont {Fodor},
  \citenamefont {Katz},\ and\ \citenamefont {Szabo}}]{Aoki:2006br}%
  \BibitemOpen
  \bibfield  {author} {\bibinfo {author} {\bibfnamefont {Y.}~\bibnamefont
  {Aoki}}, \bibinfo {author} {\bibfnamefont {Z.}~\bibnamefont {Fodor}},
  \bibinfo {author} {\bibfnamefont {S.}~\bibnamefont {Katz}}, \ and\ \bibinfo
  {author} {\bibfnamefont {K.}~\bibnamefont {Szabo}},\ }\href {\doibase
  10.1016/j.physletb.2006.10.021} {\bibfield  {journal} {\bibinfo  {journal}
  {Phys.Lett.}\ }\textbf {\bibinfo {volume} {B643}},\ \bibinfo {pages} {46}
  (\bibinfo {year} {2006}{\natexlab{a}})},\ \Eprint
  {http://arxiv.org/abs/hep-lat/0609068} {arXiv:hep-lat/0609068 [hep-lat]}
  \BibitemShut {NoStop}%
\bibitem [{\citenamefont {Aoki}\ \emph
  {et~al.}(2006{\natexlab{b}})\citenamefont {Aoki}, \citenamefont {Endrodi},
  \citenamefont {Fodor}, \citenamefont {Katz},\ and\ \citenamefont
  {Szabo}}]{Aoki:2006we}%
  \BibitemOpen
  \bibfield  {author} {\bibinfo {author} {\bibfnamefont {Y.}~\bibnamefont
  {Aoki}}, \bibinfo {author} {\bibfnamefont {G.}~\bibnamefont {Endrodi}},
  \bibinfo {author} {\bibfnamefont {Z.}~\bibnamefont {Fodor}}, \bibinfo
  {author} {\bibfnamefont {S.}~\bibnamefont {Katz}}, \ and\ \bibinfo {author}
  {\bibfnamefont {K.}~\bibnamefont {Szabo}},\ }\href {\doibase
  10.1038/nature05120} {\bibfield  {journal} {\bibinfo  {journal} {Nature}\
  }\textbf {\bibinfo {volume} {443}},\ \bibinfo {pages} {675} (\bibinfo {year}
  {2006}{\natexlab{b}})},\ \Eprint {http://arxiv.org/abs/hep-lat/0611014}
  {arXiv:hep-lat/0611014 [hep-lat]} \BibitemShut {NoStop}%
\bibitem [{\citenamefont {Bzdak}\ and\ \citenamefont
  {Skokov}(2012)}]{Bzdak:2011yy}%
  \BibitemOpen
  \bibfield  {author} {\bibinfo {author} {\bibfnamefont {A.}~\bibnamefont
  {Bzdak}}\ and\ \bibinfo {author} {\bibfnamefont {V.}~\bibnamefont {Skokov}},\
  }\href {\doibase 10.1016/j.physletb.2012.02.065} {\bibfield  {journal}
  {\bibinfo  {journal} {Phys.Lett.}\ }\textbf {\bibinfo {volume} {B710}},\
  \bibinfo {pages} {171} (\bibinfo {year} {2012})},\ \Eprint
  {http://arxiv.org/abs/1111.1949} {arXiv:1111.1949 [hep-ph]} \BibitemShut
  {NoStop}%
\bibitem [{\citenamefont {Bzdak}\ and\ \citenamefont
  {Bozek}(2013)}]{Bzdak:2013yla}%
  \BibitemOpen
  \bibfield  {author} {\bibinfo {author} {\bibfnamefont {A.}~\bibnamefont
  {Bzdak}}\ and\ \bibinfo {author} {\bibfnamefont {P.}~\bibnamefont {Bozek}},\
  }\href {\doibase 10.1016/j.physletb.2013.08.003} {\bibfield  {journal}
  {\bibinfo  {journal} {Phys.Lett.}\ }\textbf {\bibinfo {volume} {B726}},\
  \bibinfo {pages} {239} (\bibinfo {year} {2013})},\ \Eprint
  {http://arxiv.org/abs/1303.1138} {arXiv:1303.1138 [nucl-th]} \BibitemShut
  {NoStop}%
\bibitem [{\citenamefont {McLerran}\ and\ \citenamefont
  {Skokov}(2013)}]{McLerran:2013hla}%
  \BibitemOpen
  \bibfield  {author} {\bibinfo {author} {\bibfnamefont {L.}~\bibnamefont
  {McLerran}}\ and\ \bibinfo {author} {\bibfnamefont {V.}~\bibnamefont
  {Skokov}},\ }\href@noop {} {\  (\bibinfo {year} {2013})},\ \Eprint
  {http://arxiv.org/abs/1305.0774} {arXiv:1305.0774 [hep-ph]} \BibitemShut
  {NoStop}%
\bibitem [{\citenamefont {Tuchin}(2013)}]{Tuchin:2013apa}%
  \BibitemOpen
  \bibfield  {author} {\bibinfo {author} {\bibfnamefont {K.}~\bibnamefont
  {Tuchin}},\ }\href {\doibase 10.1103/PhysRevC.88.024911} {\bibfield
  {journal} {\bibinfo  {journal} {Phys.Rev.}\ }\textbf {\bibinfo {volume}
  {C88}},\ \bibinfo {pages} {024911} (\bibinfo {year} {2013})},\ \Eprint
  {http://arxiv.org/abs/1305.5806} {arXiv:1305.5806 [hep-ph]} \BibitemShut
  {NoStop}%
\bibitem [{\citenamefont {Joyce}\ and\ \citenamefont
  {Shaposhnikov}(1997)}]{Joyce:1997uy}%
  \BibitemOpen
  \bibfield  {author} {\bibinfo {author} {\bibfnamefont {M.}~\bibnamefont
  {Joyce}}\ and\ \bibinfo {author} {\bibfnamefont {M.~E.}\ \bibnamefont
  {Shaposhnikov}},\ }\href {\doibase 10.1103/PhysRevLett.79.1193} {\bibfield
  {journal} {\bibinfo  {journal} {Phys.Rev.Lett.}\ }\textbf {\bibinfo {volume}
  {79}},\ \bibinfo {pages} {1193} (\bibinfo {year} {1997})},\ \Eprint
  {http://arxiv.org/abs/astro-ph/9703005} {arXiv:astro-ph/9703005 [astro-ph]}
  \BibitemShut {NoStop}%
\bibitem [{\citenamefont {Akamatsu}\ and\ \citenamefont
  {Yamamoto}(2013)}]{Akamatsu:2013pjd}%
  \BibitemOpen
  \bibfield  {author} {\bibinfo {author} {\bibfnamefont {Y.}~\bibnamefont
  {Akamatsu}}\ and\ \bibinfo {author} {\bibfnamefont {N.}~\bibnamefont
  {Yamamoto}},\ }\href {\doibase 10.1103/PhysRevLett.111.052002} {\bibfield
  {journal} {\bibinfo  {journal} {Phys.Rev.Lett.}\ }\textbf {\bibinfo {volume}
  {111}},\ \bibinfo {pages} {052002} (\bibinfo {year} {2013})},\ \Eprint
  {http://arxiv.org/abs/1302.2125} {arXiv:1302.2125 [nucl-th]} \BibitemShut
  {NoStop}%
\end{thebibliography}%

\end{document}